\documentclass[aps,prd,twocolumn,nofootinbib,superscriptaddress]{revtex4-1} % revtex4.1 macros
\usepackage{amsmath,amssymb}
\usepackage{graphicx}
\usepackage{bigints}

\begin{document}

\title{Inflation in a closed universe}

%% Author

\author{Bharat Ratra}
\email{ratra@phys.ksu.edu}
\affiliation{Department of Physics, Kansas State University, 116 Cardwell Hall, Manhattan, Kansas 66506, USA}

\date{\today ~~~ KSUPT - 17/1}

\begin{abstract}

To derive a power spectrum for energy density inhomogeneities in a closed 
universe, we study a spatially-closed inflation-modified hot big bang model 
whose evolutionary history is divided into three epochs: an early 
slowly-rolling scalar field
inflation epoch and the usual radiation and nonrelativistic matter epochs. 
(For our purposes it is not necessary to consider a final dark energy 
dominated epoch.) We derive general solutions of the relativistic linear 
perturbation equations in each epoch. The constants of integration in the 
inflation epoch solutions are determined from de Sitter invariant 
quantum-mechanical initial conditions in the Lorentzian section of the 
inflating closed de Sitter space derived from Hawking's prescription that 
the quantum state of the universe only include field configurations that 
are regular on the Euclidean (de Sitter) sphere section. The constants of 
integration in the radiation and matter epoch solutions 
are determined from joining conditions derived by requiring that the linear 
perturbation equations remain nonsingular at the transitions between epochs.  
The matter epoch power spectrum of gauge-invariant energy density 
inhomogeneities is not a power law, and depends on spatial wavenumber in the 
way expected for a generalization to the closed model of the standard 
flat-space scale-invariant power spectrum.
The power spectrum we derive appears to differ from a number of other 
closed inflation model power spectra derived assuming different (presumably
non de Sitter invariant) initial conditions.

\end{abstract}

\maketitle
\section{Introduction}
\label{Intro}

In the standard scenario, dark energy dominates the current cosmological 
energy budget and results in the observed accelerating cosmological expansion.
Earlier on nonrelativistic (cold dark and baryonic) matter dominated, powering 
the decelerating cosmological expansion. In flat-$\Lambda$CDM 
\citep{Peebles1984}, the current ``standard'' cosmological model, Einstein's 
cosmological constant $\Lambda$ is the dark energy with nonrelativistic cold 
dark matter (CDM) being the second biggest contributor to the current energy 
budget and spatial hypersurfaces are assumed to be flat. See Refs.\ 
\citep{CosmoRev} for reviews of the dark energy picture as well as of the 
modified gravity scenario.  

The standard scenario is supported by a number of different measurements, but 
these do not rule out mildly varying --- in time and space --- dark energy or 
mildly curved spatial hypersurfaces. These measurements include cosmic 
microwave background (CMB) anisotropy observations \citep{Adeetal2016}, baryon
acoustic oscillation (BAO) data \citep{Alametal2016}, Hubble parameter versus 
redshift measurements \citep{Farooqetal2016}\footnote{
These $H(z)$ data \citep{Hzdata} are particularly interesting as they span a 
large redshift range, to almost $z = 2.4$, and show evidence consistent
with a transition from early nonrelativistic-matter-dominated deceleration 
to current dark-energy-powered acceleration \citep{Hztransition, 
Farooqetal2016}, in agreement with what is expected in $\Lambda$CDM and 
other dark energy models.}, Type Ia supernova 
apparent magnitude observations \citep{Betouleetal2014}, as well as the 
growth of structure as a function of redshift \citep{Pavlovetal2014}. 
  
Other measurements, which are not as constraining, are also consistent
with the $\Lambda$CDM model. These include HII galaxy apparent magnitude
versus redshift data \citep{HII}, galaxy cluster number counts \citep{cluster},
angular size as a function of redshift measurements \citep{AngSize}, lookback 
time observations \citep{lookback}, gamma-ray burst data \citep{GRB}, and 
cluster gas mass fraction observations \citep{gasmass}. Near-future data will
provide more restrictive and possibly very interesting constraints 
\citep{nearfuture}.

It is reassuring that most current measurements are not inconsistent with 
the standard flat-$\Lambda$CDM model, although they are also not inconsistent 
with weakly varying dark energy or a mild amount of space curvature. To be 
able to distinguish between the options and better pin down cosmological 
parameter values will require resolution of a number of issues. For instance,
for over a decade and a half now, median statistics analyses of Huchra's 
growing compilation of Hubble constant $H_0$ measurements have been 
consistent with $H_0 = 68 \pm 2.8$ km s$^{-1}$ Mpc$^{-1}$ 
\citep{ChenRatra2011}, in good agreement with the range of values recently 
estimated from CMB anisotropy data \citep{CMBH0, Adeetal2016}, BAO 
observations \citep{BAOH0, Alametal2016}, Hubble parameter measurements 
\citep{HzH0}, and from a compilation of recent
cosmological data and the standard model of particle physics with only 
three light neutrino species \citep{Calabreseetal2012}. Unfortunately, 
however, local measurements of the expansion rate favor a significantly 
larger value, $H_0 = 73.24 \pm 1.74$ km s$^{-1}$ Mpc$^{-1}$ 
\citep{Riessetal2016}, larger than what is favored by a number of other 
observations. Until this difference is understood and resolved, it is 
probably wiser to proceed cautiously about judging the viability of 
cosmological 
models.\footnote{Of course, similar issues affect measurements of other 
parameters.}  

That said, there have been a number of recent papers suggesting that 
the predictions of flat-$\Lambda$CDM might not be compatible with some 
$H(z)$ data \citep{Hzdynamical}, as well as with a combination of 
cosmological observations \citep{Solaetal, Zhangetal2017}, and that 
dynamical dark energy 
provides a better fit to these measurements. If this is supported by more 
and better-quality data, it will be an important clue about the nature of 
the dark energy. On the other hand, it would be useful to check if 
these data were also in accord with a non-flat $\Lambda$CDM model or if
they prefer dynamical dark energy over spatial curvature.

Compared to the time-independent cosmological constant, a time-varying 
dark energy density evolves in a manner closer to that of spatial curvature
energy density and this can cause a complication. For instance, when CMB 
anisotropy measurements are studied in the context of the $\Lambda$CDM model,
they indicate that spatial hypersurfaces are close to flat, although a mild 
amount of curvature is still allowed \citep{Adeetal2016}.
On the other hand, under the assumption of flat spatial geometry these 
measurements favor a time-independent dark energy density, although mild
dark energy time evolution remains an option. However, if CMB
anisotropy data are analyzed using a non-flat dynamical dark energy 
model, there is degeneracy between space curvature and the parameter that 
governs the dark energy density, resulting in weaker constraints on both 
parameters when compared to the case when only either dark energy density 
time variability or non-zero spatial curvature is assumed 
\citep{timevaryingnonflatconstraints}. This is the case for other data 
also, see Refs.\ \citep{Farooqetal2016, curvlimit}.\footnote{
See Ref.\ \citep{Leonardetal2016} for potential constraints on space 
curvature from proposed experiments.}     

The simplest physically-consistent dynamical dark energy model is $\phi$CDM
\citep{PeeblesRatra1988, RatraPeebles1988}.\footnote{
While the XCDM parameterization is often used to model dynamical dark 
energy, it is an incomplete and inconsistent parameterization (as it cannot 
describe inhomogeneities). It also does not accurately model even the 
spatially homogeneous part of the $\phi$CDM model \citep{PodariuRatra2000}.}
Here dark energy is a scalar field $\phi$ with a potential energy density 
$V(\phi)$ that gradually decreases 
with increasing $\phi$. The original $\phi$CDM model assumed flat spatial 
hypersurfaces.This was generalized to the non-flat case in Ref.\ 
\citep{Pavlovetal2013}; the time-dependent attractor solution discovered 
in the spatially-flat case is also present in the non-flat case.

To complete this non-flat dynamical dark energy model requires a prescription 
for what happens at very early times in the model. This is provided by 
inflation, \citep{InflationOriginal, Martinetal2014}, which is easily 
generalized to the 
spatially-open case in the Gott open-bubble inflation model 
\citep{OpenInflation}. In this model a spatially-open bubble nucleates and 
then inflates only for a limited time so spatial curvature is not 
completely diluted. If necessary, an earlier epoch 
of less-limited inflation can be used to explain spatial homogeneity.\footnote{
Alternately, if the bubble nucleation process is slow enough it might be 
possible to arrange for the interior of the open bubble to be homogeneous 
enough.}

In this initial hyperbolic (or open) de Sitter space of the open bubble, the 
standard requirement that the ground state energy of the (appropriately 
rescaled) scalar inflaton field spatial inhomogeneity not diverge in the
scale factor $a \rightarrow 0$ limit provides the needed initial condition
\citep{Ratra1994} and results in a late-time energy density inhomogeneity 
power spectrum \citep{RatraPeebles1994, RatraPeebles1995} that is the 
generalization to the open case \citep{LythStewart1990} of the 
scale-invariant spectrum of the flat model \citep{HPYZ}.

Perhaps the simplest model of inflation in a closed universe is that based 
on Hawking's prescription for the quantum state of the universe 
\citep{Hawking1984}. Hawking proposes including in the functional integral 
only those field configurations which are regular on the Euclidean section
\citep{Hawking1984, Ratra1985}. This may be viewed as the nucleation of a 
closed de Sitter Lanczos universe on the Lorentzian section, because the waist
of the Lorentzian de Sitter Lanczos hyperboloid and the equator of the 
Euclidean (de Sitter Lanczos) sphere are identified \citep{Hawking1984, 
Ratra1985}. For variants of this 
scenario see Refs.\ \citep{Linde, Grattonetal2002, LasenbyDoran2005}. If the 
nucleation process is slow enough it might be possible to make the nucleated
Lorentzian closed de Sitter space sufficiently spatially homogeneous. See 
Refs.\ \citep{Ellis} for discussions of homogeneity in a more conventional 
closed inflation model.

During the Lorentzian closed de Sitter expansion, quantum mechanical spatial 
inhomogeneities 
in the scalar inflaton field could provide the needed density inhomogeneities.
A major advantage of the Hawking proposal is that it provides reasonable 
quantum mechanical initial conditions for these fluctuations. In the closed
de Sitter model the $a \rightarrow 0$ limit does not lie in the Lorentzian 
section \citep{Ratra1985}, unlike in the open and flat cases. Remarkably, 
Hawking's prescription of only including field configurations regular on the 
Euclidean section does in fact correspond to the ground state energy of the 
(appropriately rescaled) scalar field inhomogeneity not diverging as 
$a \rightarrow 0$, which in this case is either the north or south pole of the 
Euclidean section sphere (actually there are an infinite number of spheres, 
each connected to the next at the poles) \citep{Ratra1985}, and in fact 
leads to a de Sitter invariant ground state scalar field two-point correlation 
function, \citep{Ratra1985}. It is likely that this is the unique initial 
condition with this 
property \citep{Ratra1985}.  
    
In this paper we use this initial condition to compute the energy density 
inhomogeneity power spectrum in a closed universe in terms of the potential 
of the inflaton and other parameters. The spatial wavenumber dependence 
of the late-time spectrum we find, using a simple inflation model, is the 
generalization of the scale-invariant spectrum in the spatially-flat case 
\citep{HPYZ} to the closed universe \citep{WhiteScott1996, closedPk, 
Grattonetal2002}.\footnote{See Ref.\ \citep{PChenetal2017} for a discussion
of a massive scalar field inflaton closed inflation model.} 
There have also been a number of earlier computations of spectra in the 
closed model \citep{LasenbyDoran2005, Massoetal2008, Bongaetal2016}, using 
different initial conditions compared to what we have used here
(also see Ref.\ \citep{Efstathiou2003}). We emphasize that the initial 
conditions we use here results in a scalar 
field two-point function that is de Sitter invariant \citep{Ratra1985} 
and it is unclear how to interpret any other initial condition. 

A proper analysis of CMB anisotropy data in a slightly closed model --- 
which is consistent with current observations --- will make use of the 
spectrum we have derived here. While all that is needed for such an analysis 
is the spectral shape of the power spectrum (not the overall amplitude), 
which was previously known, it is also important to show that a 
computation using Hawking's initial conditions in a consistent inflation model 
--- as done here --- does result in such a power spectrum. We have also 
established that the de Sitter invariant initial conditions \citep{Hawking1984,
Ratra1985} do result in the expected power spectrum \citep{WhiteScott1996}
that differs from those
found in Refs.\ \citep{LasenbyDoran2005, Massoetal2008, Bongaetal2016}.

It seems that flat-$\Lambda$CDM, which is consistent with most observations, 
predicts more large-angle (low multipole $\ell$) CMB temperature anisotropy 
power than is observed \citep{Adeetal2016}. In the context of inflation 
and the energy density inhomogeneity power spectrum derived here and in the 
open inflation model \citep{OpenInflation}, going to a slightly non-flat 
(closed) $\Lambda$CDM (or dynamical dark energy) model might help reduce this 
low-$\ell$ discrepancy \citep{Oobaetal2017}, also see Ref.\ 
\citep{Oobaetal2017b}.

In Sec.\ II we review the spatial geometry of the closed model and various 
properties of the eigenfunctions of the spatial Laplacian. Synchronous gauge 
linear perturbation equations, in both the scalar field inflation epoch and 
fluid (radiation and nonrelativistic matter) epochs, are derived in 
Sec.\ III, where we also list the scalar (under general coordinate 
transformations) parts of these equations in spatial momentum space. These 
are used to establish that the synchronous gauge linear perturbation 
equations of a fluid model with a specified spacetime-dependent `speed of 
sound' coincide with those of the scalar field model (a generalization of 
the flat model result of Ref.\  \citep{Ratra1991a}). In Sec.\ III D, we 
examine how the (scalar) synchronous gauge variables transform under the 
remnants of general coordinate invariance, construct gauge-invariant 
combinations of these variables, and derive equations of motion for these 
gauge-invariant variables. In Sec.\ IV we solve the inflation epoch 
equations and determine the constants of integration in the general solution 
for the perturbations by using the Hawking initial conditions. Here we also 
list expressions for the gauge-invariant variables, and compute the scalar
field and energy density perturbation two-point correlation functions. 
In Sec.\ V A we derive general solutions for the gauge-invariant variables 
in the radiation epoch; in Sec.\ V B we solve the synchronous gauge 
equations in the nonrelativistic matter epoch, and list expressions for 
the gauge-invariant variables in this epoch. The general solutions in the 
radiation and matter epochs depend on constants of integration which are 
determined from joining conditions derived by requiring that the equations 
of motion be nonsingular at the transitions; these are listed in Sec.\ VI A. 
The constants of integration are determined in Sec.\ VI B while in Sec.\ 
VI C we extract the large-scale contribution to these expressions for the 
constants. Nonrelativistic matter epoch theoretical expressions characterizing 
large-scale structure are most conveniently compared to observational data 
on a spatial hypersurface on which the time derivative of the trace of the 
metric perturbation has been set to zero --- this is the instantaneously 
Newtonian spatial hypersurface.  We construct these coordinates, and 
list expressions for the relevant power spectra, in Sec.\ VII, where we
also record the gauge-invariant energy density inhomogeneity power spectrum.
We conclude in Sec.\ VIII.

\section{Technical Preliminaries}
\label{technicalpreliminaries}

The positive spatial curvature (closed) FLRW model has the line element
\begin{eqnarray}
\label{lineelement}
    ds^2 & = & dt^2 - a^2(t) H_{ij} (\vec x) dx^i \, dx^j \\ 
         & = & dt^2 - a^2(t) \left [ d\chi^2
          + {\rm sin}^2\!(\chi) \left\{d\theta^2 + {\rm sin}^2\!
          (\theta ) \, d\phi^2 \right\} \right], \nonumber
\end{eqnarray}
where $a(t)$ is the FLRW scale factor, $H_{ij}(\vec x)$ the metric on
the closed spatial hypersurface, the `radial' coordinate $0 \le \chi < 
\pi$, and $\theta , \phi$ are the usual angular coordinates on the two-sphere. 
The square of the distance between two points, $(t, \chi , \theta , \phi )$
and $(t, \chi' , \theta' , \phi' )$, is
\begin{eqnarray}
   \sigma^2  =  2 a^2 (t) 
              \left[ - 1 + {\rm cos}(\gamma_3) \right],
\end{eqnarray}
\begin{eqnarray}
\label{gamma3}
   {\rm cos}(\gamma_3)  =  {\rm cos}(\chi) {\rm cos}(\chi')
     + {\rm sin}(\chi) {\rm sin}(\chi') {\rm cos}(\gamma_2),
     \cr
\end{eqnarray}
where $\gamma_2$ is the usual angle between the two points $(\theta , \phi)$
and $(\theta' , \phi')$ on the two-sphere
\begin{eqnarray}
   {\rm cos}(\gamma_2) = {\rm cos}(\theta) {\rm cos}(\theta')
     + {\rm sin}(\theta) {\rm sin}(\theta') {\rm cos}(\phi - \phi').
\end{eqnarray}

The three-dimensional spatial covariant derivative of a spatial
vector (or tensor) will be denoted by a ${ }_{|}$, is defined 
in the usual way
\begin{eqnarray}
      A^i{}_{|j} & = & A^i{}_{,j} + \Gamma^i{}_{jk} A^k ,  \nonumber \\
      A_{i|j} & = & A_{i,j} - \Gamma^k{}_{ij} A_k ,
\end{eqnarray}
where the commas denote spatial differentiation, and obeys the usual
relations of covariant differentiation. The three-dimensional 
spatial Christoffel symbol is
\begin{eqnarray}
   \Gamma^i{}_{jk} = {1 \over 2} H^{il} \left ( H_{lj,k} + H_{lk,j}
         - H_{jk,l} \right) .
\end{eqnarray}
The ${}_|$ operator obeys the usual relations of covariant differentiation.

The three-dimensional spatial Laplacian for the metric of Eq.\ 
(\ref{lineelement}) is
\begin{widetext}
\begin{eqnarray}
     L^2 = {1 \over {\rm sin}^2\!(\chi)}
                   {\partial \over \partial\chi}
                   \left( {\rm sin}^2\!(\chi)
                   {\partial \over \partial\chi} \right) 
           + {1 \over {\rm sin}^2\!(\chi) {\rm sin}(\theta)}
                   {\partial \over \partial\theta}
                   \left( {\rm sin} (\theta) {\partial \over \partial \theta}
                   \right) 
           + {1 \over {\rm sin}^2\!(\chi) {\rm sin}^2\! (\theta)}
                   {\partial^2 \over \partial\phi^2} .
\end{eqnarray}  
The scalar eigenfunctions
$Y_{ABC}$ of $L^2$ obey, \citep{Harmonics, Ratra1985},
\begin{eqnarray}
\label{eigenvalue}
   L^2 Y_{ABC} (\Omega) =  H^{ij}(\Omega) \left[Y_{ABC} (\Omega) \right]
     _{|i|j}  =  - A ( A + 2 ) Y_{ABC} (\Omega) , 
\end{eqnarray}
where $\Omega = (\chi , \theta , \phi)$, integer $A = 0, 1, 2 \cdots$,
and the two `magnetic' integral indices 
$B\epsilon [-A, A],$ and $C\epsilon [-B, B]$. {\bf The $O(4)$ symmetry 
makes the spatial Laplacian eigenvalues independent of the two 
magnetic indices $B$ and $C$, see discussion in App.\ B of Ref.\ 
\citep{Ratra1985}.} 
The orthonormal eigenfunctions are, \citep{Harmonics, Ratra1985},
\begin{eqnarray}
   Y_{ABC} (\Omega)  =   \sqrt{(A + 1) \Gamma\left(A + B + 2\right)
        \over \Gamma \left(A - B + 1 \right) }
        \left[ {\rm sin} (\chi) 
        \right]^{-1/2} P^{-B -1/2}_{A -1/2} \left( {\rm cos}
        (\chi)\right) Y_{BC} (\theta , \phi) ,
\end{eqnarray}
where $Y_{BC}$ is the standard two-dimensional spherical 
harmonic, $\Gamma$ is the gamma
function, and $P^\mu_\nu$ is the associated Legendre function of
the first kind (Chap.\ 3 of Ref.\ \citep{EHTF} or Chap.\ 8 of Ref.\ 
\citep{AS}). The orthonormality relation is
\begin{eqnarray}
    \int^\pi_0 d\chi \, {\rm sin}^2\! (\chi)  
    \int_{S^2} d\Omega_2 \, Y_{ABC}(\Omega)
    \left[Y_{A'B'C'}(\Omega)\right]^* 
    = \delta_{A,A'} \delta_{B,B'}
    \delta_{C,C'} ,
\end{eqnarray}
where $S^2$ is the two-dimensional unit sphere with volume element 
$d\Omega_2$, and $\delta_{A,A'}, \delta_{B,B'},$ and $\delta_{C,C'}$
are Kronecker deltas. The addition theorem is, \citep{Ratra1985},
\begin{eqnarray}
   P^{-1/2}_{A + 1/2} \left({\rm cos}(\gamma_3)\right) 
   = {(2\pi)^{3/2}\over (A + 1)^2}
     \left[{\rm sin} (\gamma_3)\right]^{1/2} 
     \sum_{B,C} Y_{ABC}(\Omega)
     \left[Y_{ABC}(\Omega')\right]^* ,
\end{eqnarray}
\end{widetext}
where $\gamma_3$ is in Eq.\ (\ref{gamma3}).

We shall have need for the following relations, which may be derived by using 
standard manipulations (see the first of Refs.\ \citep{Harmonics}),
\begin{eqnarray}
      Y_{|i|j} & = & Y_{|j|i} , \\
      H^{jk} Y_{|k|i|j} & = & - (A^2 + 2A - 2) Y_{|i} , \\
      H^{kl} Y_{|l|j|i|k} & = & - (A^2 + 2A - 5) Y_{|i|j} \nonumber \\
                          & { } & + A(A + 2) Y H_{ij} , \\
      H^{kl} Y_{|i|j|k|l} & = & - (A^2 + 2A -6) Y_{|i|j} \nonumber \\ 
                          & { } & + 2 A(A + 2) Y H_{ij},
\end{eqnarray}
where we have suppressed the spatial momentum indices on the scalar
(under the spatial reparameterization remnants of general coordinate 
transformations in synchronous gauge) spatial harmonic $Y_{ABC}
(\Omega)$. 

Also, the Ricci tensor on the spatial hypersurface is
\begin{eqnarray}
   {}^{(3)}R_{ij} = \Gamma^k{}_{ij,k} - \Gamma^k{}_{ki,j}
         + \Gamma^k{}_{kl} \Gamma^l{}_{ij} - \Gamma^k{}_{lj}
         \Gamma^l{}_{ki} ,
\end{eqnarray}
and it may be shown that for the spatial metric given in Eq.\ 
(\ref{lineelement}),
\begin{eqnarray}
   {}^{(3)}R_{ij} =  2 H_{ij} . 
\end{eqnarray}

\section{Equations of Motion}
\label{equationsofmotion}

In this section we derive the general, closed FLRW model, position
space, synchronous gauge, linear perturbation theory equations of motion, 
for both the homogeneous background fields and for the spatial 
irregularities, in the early time scalar field inflation
epoch and in the late time ideal fluid (radiation or matter) epochs.
(The current dark energy dominated epoch is not as analytically tractable 
and so is ignored here; our matter epoch results suffice for our 
purposes.)  
We then extract the scalar (under general coordinate transformations) parts
of these equations (i.e., we ignore transverse peculiar velocity 
perturbations and gravitational wave perturbations), and record
their spatial momentum space form.

For later use, we establish that the synchronous gauge linear 
perturbation theory equations of a fluid model which allows for a 
specified spacetime-dependent `speed of sound' are identical to the scalar 
field model synchronous gauge linear perturbation equations.

We also examine how the (scalar) synchronous gauge spatial irregularity
variables of interest transform under the remnants of general coordinate
invariance in synchronous gauge, write down combinations of these 
variables that are invariant under these transformations, and derive the 
equations of motion for these gauge-invariant variables.

\subsection{Einstein-scalar-field model equations of motion}
\label{einsteinscalarfield}

The Einstein-scalar field action, for the metric tensor $g_{\mu\nu}$
and inflaton scalar field $\Phi$, is
\begin{eqnarray}
   & { } &  S  = \\
   & { } & {m_p{}^2 \over 16\pi} \int dt \, d^3\! x 
           \sqrt{-g} \left[ -R + {1 \over 2}
           g^{\mu\nu} \partial_\mu\Phi \partial_\nu\Phi -
           {1 \over 2} V(\Phi) \right] , \nonumber 
\end{eqnarray}
where $m_p = G^{-1/2}$ is the Planck mass. Varying, we find the equations of 
motion,
\begin{eqnarray}
\label{scalareom}
    & {} & {1\over \sqrt {-g}} \partial_\mu \left( \sqrt {-g} g^{\mu\nu}
      \partial_\nu \Phi \right) + {1 \over 2} V'(\Phi) = 0 ,
\end{eqnarray}
\begin{eqnarray}
\label{einsteinscalareom}
    & {} & R_{\mu\nu} = {8\pi \over m_p{}^2} \left( T_{\mu\nu}
      - {1\over 2} g_{\mu\nu} T \right) ,
\end{eqnarray}
where prime denotes a derivative with respect to $\Phi$ and 
$T$ is the trace of the stress-energy tensor,
\begin{eqnarray}
\label{scalarstresstensor}
   & { } & T_{\mu\nu} = \nonumber \\
   & { } & {m_p{}^2 \over 16\pi} \left[ \partial_\mu \Phi
           \partial_\nu\Phi - {1\over 2} g_{\mu\nu} \left\{
           g^{\lambda\rho} \partial_\lambda\Phi \partial_\rho\Phi
           - V(\Phi) \right\}\right] .
\end{eqnarray}

To derive the equations of motion for the spatially homogeneous background
fields and for the spatial irregularities, we linearize eqs. (\ref{scalareom}) 
-- (\ref{scalarstresstensor}) about a closed FLRW model and a spatially 
homogeneous scalar field. We work in synchronous gauge, with line element
\begin{eqnarray}
\label{pertlineelement}
   ds^2 = dt^2 - a^2(t) \left[ H_{ij} (\vec x) - h_{ij} (t, \vec x)
          \right] dx^i dx^j ,
\end{eqnarray}
where the background metric on the closed spatial hypersurfaces, $H_{ij}$,
is given in eq. (\ref{lineelement}), and the metric perturbations are 
denoted by $h_{ij}$. The expansion for the scalar field is
\begin{eqnarray}
   \Phi (t, \vec x) = \Phi_b (t) + \phi (t, \vec x) ,
\end{eqnarray}
where $\Phi_b$ and $\phi$ are the spatially homogeneous and 
inhomogeneous parts of the scalar field (the scalar field perturbation
$\phi$ should not be confused with the angular variable $\phi$ of
Sec.\ \ref{technicalpreliminaries}). The linearized stress-energy tensor 
components are
\begin{widetext}
\begin{eqnarray}
\label{scalarT00}
      T_{00} = {m_p{}^2 \over 32\pi} \left[\dot\Phi_b{}^2
                 + V(\Phi_b) \right] 
             + {m_p{}^2 \over 16\pi}
                 \left[ \dot\Phi_b\dot\phi + {1\over 2} V'(\Phi_b)
                 \phi \right] + \cdots ,
\end{eqnarray}
\begin{eqnarray}
\label{scalarT0i}
      T_{0i}  =  {m_p{}^2 \over 16\pi} \dot\Phi_b \partial_i\phi
                 + \cdots ,
\end{eqnarray}
\begin{eqnarray}
\label{scalarTij}
      T_{ij} =  {m_p{}^2 \over 32 \pi} a^2 H_{ij} \left[ \dot\Phi_b{}^2
                 - V(\Phi_b) \right] 
             + {m_p{}^2 \over 16\pi} a^2
                 \bigg[ H_{ij} \left\{ \dot\Phi_b\dot\phi - 
                 {1 \over 2} V'(\Phi_b) \phi \right\} 
             & {} & \ \ - {1\over 2} h_{ij} \left\{ \dot\Phi_b{}^2
                 - V(\Phi_b) \right\} \bigg] + \cdots ,
\end{eqnarray}
\end{widetext}
where the ellipses denote terms of second and higher order in the 
perturbations.

The equations of motion for the spatially homogeneous parts of the fields are
\begin{eqnarray}
\label{scalarfieldKG}
   \ddot\Phi_b + 3 {\dot a \over a} \dot\Phi_b + {1\over 2} 
                   V'(\Phi_b) = 0 ,
\end{eqnarray}
\begin{eqnarray}
\label{scalarfieldfriedmann1}
   \left({\dot a \over a}\right)^2 = {1\over 12} \left[ \dot
                   \Phi_b{}^2 + V(\Phi_b) \right] - {1 \over a^2} ,
\end{eqnarray}
\begin{eqnarray}
\label{scalarfieldfriedmann2}                   
   {\ddot a \over a } = - {1\over 6} \dot\Phi_b{}^2 + 
                    {1\over 12} V(\Phi_b) , 
\end{eqnarray}
where an overdot denotes a derivative with respect to time. 
The only change, relative to the equations for the flat model 
(Sec.\ VII of Ref.\ \citep{RatraPeebles1988} and Sec.\ II of 
Ref.\ \citep{Ratra1991a}), is the new 
term $(1/a^2)$ on the right hand side of eq.\ (\ref{scalarfieldfriedmann1}). 
The first order perturbation equations are
\begin{eqnarray}
\label{scalarpert1}
\ddot\phi + 3 {\dot a \over a} \dot\phi - {L^2 \over a^2}
                    \phi + {1\over 2} V''(\Phi_b) \phi =
                    {1\over 2} \dot h \dot \Phi_b ,
\end{eqnarray}
\begin{eqnarray}
\label{scalarpert2}
\ddot h + 2 {\dot a \over a} \dot h = 2 \dot\Phi_b \dot \phi
                    - {1\over 2} V'(\Phi_b) \phi ,
\end{eqnarray}
\begin{eqnarray}
\label{scalarpert3}
\dot h_{|i} - \left( H^{jk}\dot h_{ki}\right)_{|j} =
                    \dot\Phi_b \phi_{|i} ,
\end{eqnarray}
\begin{eqnarray}
\label{scalarpert4}
\ddot h_{ij} & + & 3 {\dot a \over a} \dot h_{ij}
                    + {\dot a \over a} H_{ij} \dot h
                    - {1\over a^2} h_{|i|j} \nonumber \\ 
             & + & {1\over a^2} \left[ H^{kl} \left(
                    h_{li|j} + h_{lj|i}  
                    - h_{ij|l} \right) \right]_{|k} - 
                    {4 \over a^2} h_{ij} \nonumber \\ 
             & = & - {1\over 2} H_{ij}
                    V'(\Phi_b) \phi ,
\end{eqnarray}
where the trace of the metric perturbation is denoted by $h \, (=
H^{ij} h_{ij})$ and spatial indices are raised and lowered with
the background metric $H_{ij}$. Eq.\ (\ref{scalarpert1}) governs the 
evolution of the scalar field perturbation, eq.\ (\ref{scalarpert2}) 
that of the trace of the metric perturbation, and eqs.\ 
(\ref{scalarpert3}) and (\ref{scalarpert4}) that of the remaining part 
of the metric perturbation. Besides the expected change, relative to 
the equations of the flat model (Sec.\ VII of Ref.\ \citep{RatraPeebles1988} 
and Sec.\ II of Ref.\ \citep{Ratra1991a}), of all spatial derivatives being
replaced by spatial covariant derivatives, the only other change is 
the new last term on the left hand side of eq.\ (\ref{scalarpert4}), 
$4h_{ij}/a^2$.

To extract the scalar parts of eqs. (\ref{scalarpert1}) -- (\ref{scalarpert4}) 
in spatial momentum space we focus on a mode with spatial momentum 
characterized by the indices $(A, B, C)$, \citep{Harmonics},
\begin{eqnarray}
\label{phidecomposition}
     \phi (\Omega, t) = & \phi (A, B, C, t) Y (\Omega) , 
\end{eqnarray}
\begin{eqnarray}
\label{hijdecomposition}
     h_{ij} (\Omega, t) & = & {1\over 3} h(A, B, C, t) H_{ij} (\Omega)
                               Y (\Omega) \\
                        & {} & + {\cal H} (A, B, C, t) \left[ {Y_{|i|j} 
                               (\Omega) \over A(A + 2)} + {1\over 3} 
                               H_{ij} (\Omega) Y (\Omega) \right] , \nonumber 
\end{eqnarray}
where $h(A, B, C, t)$ is the trace of the metric perturbation (the 
perturbation to the size of the proper volume element) and ${\cal H}
(A, B, C, t)$ is the trace-free part (the shearing perturbation of 
the volume element). Eq.\ (\ref{hijdecomposition}) is the most general 
decomposition of the scalar part of the metric perturbation (we 
have ignored gravitational wave perturbations). The scalar parts of eqs.\ 
(\ref{scalarpert1}) -- (\ref{scalarpert4}) for a given mode in spatial 
momentum space are
\begin{eqnarray}
\label{scalarpert1s}
     \ddot\phi + 3 {\dot a \over a} \dot\phi + {A(A + 2) \over a^2}
                   \phi + {1\over 2} V''(\Phi_b) \phi = {1\over 2}
                   \dot h \dot\Phi_b , 
\end{eqnarray}
\begin{eqnarray}
\label{scalarpert2s}
     \ddot h + 2 {\dot a \over a} \dot h = 2 \dot\Phi_b\dot\phi -
                   {1\over 2} V'(\Phi_b) \phi ,
\end{eqnarray}
\begin{eqnarray}
\label{scalarpert3s}
     \dot{\cal H} = {A(A + 2) \over (A - 1) (A + 3)} 
                   \left[ {3\over 2} \dot\Phi_b\phi - \dot h \right] ,
\end{eqnarray}
\begin{eqnarray}
\label{scalarpert4s}
     \ddot h + 6 {\dot a \over a} \dot h 
                & + & {(A^2 + 2 A -4) \over a^2} h + \ddot {\cal H} 
                      + 3 {\dot a \over a} \dot {\cal H} \nonumber \\
                & + & {(A^2 + 2 A - 4) \over a^2} {\cal H} 
                       = - {3\over 2} V'(\Phi_b) \phi ,
\end{eqnarray}
\begin{eqnarray}
\label{scalarpert5s}
     \ddot {\cal H} + 3 {\dot a \over a} \dot {\cal H} 
                 - {A (A + 2) \over 3 a^2} {\cal H} 
                 - {A(A + 2) \over 3 a^2} h = 0 . 
\end{eqnarray}

\subsection{Einstein-fluid model equations of motion}
\label{einsteinfluid}

The fluid model equations of motion are covariant conservation of stress-energy
\begin{eqnarray}
\label{stressenergyconservation}
    T_\alpha{}^\beta{}_{;\beta} = 0 ,
\end{eqnarray}
and the Einstein equations, eq.\ (\ref{einsteinscalareom}), where the 
stress-energy tensor for the fluid is
\begin{eqnarray}
\label{fluidstresstensor}
    T^{\mu\nu} = (\rho + p ) u^\mu u^\nu - g^{\mu\nu} p , 
\end{eqnarray}
where $\rho$ and $p$ are the fluid energy density and pressure
and $u^\mu$ is the fluid coordinate peculiar velocity.

To derive the equations of motion for the spatially homogeneous 
background fields and for the spatial irregularities, we linearize
eqs.\ (\ref{stressenergyconservation}), (\ref{einsteinscalareom}) and 
(\ref{fluidstresstensor}) about a spatially closed FLRW model
and a spatially homogeneous background fluid. We work in synchronous 
gauge, with the line-element of eq.\ (\ref{pertlineelement}). The 
expansions for the fluid variables are
\begin{eqnarray}
     \rho (t, \vec x) = \rho_b (t) [ 1 + \delta (t, \vec x) ] ,
\end{eqnarray}
\begin{eqnarray}
     p (t, \vec x) = p_b (t) + c_s{}^2 \rho_b (t) \delta (t, \vec x) ,
\end{eqnarray}
\begin{eqnarray}
     u^0 (t, \vec x) = 1 , 
\end{eqnarray}
\begin{eqnarray}
     u^i (t, \vec x) = 0 + u^i (t, \vec x) , 
\end{eqnarray}
i.e., $u^i$ is taken to be of the same order as the fractional 
perturbation in the fluid energy density, $\delta$. Here $\rho_b$
and $p_b$ are the homogeneous background fluid energy density and 
pressure and the background equation of state is taken to be
\begin{eqnarray}
\label{eos}
    p_b (t) = \nu \rho_b (t) ,
\end{eqnarray}
where $\nu$ is a constant. The speed of propagation of `acoustic'
waves is
\begin{eqnarray}
    c_s{}^2 = {dp \over d\rho } ,
\end{eqnarray}
and, for the present, will be allowed to be a function of the spacetime 
coordinates. Expanding the fluid stress-energy tensor, eq.\ 
(\ref{fluidstresstensor}), we find 
the components
\begin{eqnarray}
\label{fluidT00}
     T_{00} = \rho_b + \rho_b \delta + \cdots ,
\end{eqnarray}
\begin{eqnarray}
\label{fluidT0i}
     T_{0i} & = & - a^2 \left(\rho_b + p_b\right)
                H_{ij} u^j + \cdots ,
\end{eqnarray}
\begin{eqnarray}
\label{fluidTij}
     T_{ij} & = & a^2 H_{ij} p_b + a^2 \left( c_s{}^2 \rho_b \delta H_{ij}
                - p_b h_{ij} \right) + \cdots , 
\end{eqnarray}
where the ellipses denote terms of quadratic and higher order in the 
perturbations.

The equations of motion for the 
spatially homogeneous parts of the fields are
\begin{eqnarray}
\label{fluiddensity}
      \dot \rho_b = - 3 {\dot a \over a} \left(\rho_b + p_b\right) ,
\end{eqnarray}
\begin{eqnarray}
\label{fluidfriedmann1}
      \left({\dot a \over a}\right)^2 = {8\pi \over 3m_p{}^2} \rho_b
                      - {1\over a^2} ,
\end{eqnarray}
\begin{eqnarray}
\label{fluidfriedmann2}
      {\ddot a \over a} = - {4\pi \over 3m_p{}^2} \left( \rho_b 
                      + 3 p_b\right) .
\end{eqnarray}
The only change, relative to the equations of the flat model (Secs.\ 82 and 85 
of Ref.\ \citep{LSSU} and Sec.\ I of Ref.\ \citep{Ratra1988}), is the new term 
$(1/a^2)$ on the right hand side of eq.\ (\ref{fluidfriedmann1}). 
The first order perturbation equations are
\begin{eqnarray}
\label{fluidpert1}
       \rho_b \dot\delta - \left( \rho_b + p_b \right) \left( {1\over 2}
          \dot h - u^i{}_{|i} \right) = 3 {\dot a \over a} \left(
          p_b - c_s{}^2\rho_b \right) \delta ,
\end{eqnarray}
\begin{eqnarray}
\label{fluidpert2}
       \ddot h + 2 {\dot a \over a} \dot h = {8 \pi \over m_p{}^2}
          \left( 1 + 3 c_s{}^2 \right) \rho_b \delta , 
\end{eqnarray}
\begin{eqnarray}
\label{fluidpert3}
      \left[ a^5 \left( \rho_b + p_b \right) H_{kl} u^l \right]_{,0} 
       = - a^3 \left( c_s{}^2 \rho_b \delta \right)_{|k} ,
 \end{eqnarray}
\begin{eqnarray}
\label{fluidpert4}
       \dot h_{|i} - \left( H^{jk} \dot h_{ki} \right)_{|j}
          = - {16\pi \over m_p{}^2} a^2 \left( \rho_b + p_b\right)
          H_{ij} u^j ,
\end{eqnarray}
\begin{eqnarray}
\label{fluidpert5}
      \ddot h_{ij} 
      & + & 3 {\dot a \over a} \dot h_{ij} + {\dot a \over a}
            H_{ij} \dot h - {1\over a^2} h_{|i|j} 
            + {1\over a^2} \bigg[ H^{kl} \big( h_{li|j} + h_{lj|i} \nonumber \\
      & - & h_{ij|l} \big) \bigg]_{|k} - {4 \over a^2} h_{ij} 
            = - {8\pi \over m_p{}^2} H_{ij} \left( 1 - c_s{}^2\right)\rho_b
            \delta . 
\end{eqnarray}
Eq.\ (\ref{fluidpert1}) governs the evolution of the fractional energy density 
perturbation, eq.\ (\ref{fluidpert3}) that of the peculiar velocity 
perturbation, eq.\ (\ref{fluidpert2}) that of the trace of the metric 
perturbation, and eqs.\ (\ref{fluidpert4}) and (\ref{fluidpert5}) that of the 
remaining part of the metric perturbation. Besides the expected change 
relative to the equations of the flat model (Sec.\ II of Ref.\ 
\citep{Ratra1991a}), of all spatial derivatives being replaced by spatial 
covariant derivatives, the only other change is the new last term on the 
left hand side of eq.\ (\ref{fluidpert5}), $4h_{ij}/a^2$.

To extract the scalar parts of eqs.\ (\ref{fluidpert1}) -- 
(\ref{fluidpert5}) in spatial momentum space we focus on a mode with 
spatial momentum characterized by the indices $(A, B, C)$ and write
\begin{eqnarray}
\label{deltadecomposition}
    \delta (\Omega , t) = \delta (A, B, C, t) Y(\Omega) ,
\end{eqnarray}
\begin{eqnarray}
\label{videcomposition}
    u_i (\Omega , t) = - {1 \over A(A + 2)} u(A, B, C, t) Y_{|i}
                               (\Omega) ;
\end{eqnarray}
we also use the metric perturbation decomposition of eq.\ 
(\ref{hijdecomposition}). Eq.\ (\ref{videcomposition}) only accounts for 
longitudinal peculiar velocity perturbations (we ignore the transverse 
peculiar velocity). The scalar parts of eqs.\ (\ref{fluidpert1}) -- 
(\ref{fluidpert5}), for a given mode, are
\begin{eqnarray}
\label{fluidpert1s}
       \rho_b \dot\delta - \left( \rho_b + p_b \right) \left( {1\over 2} 
                   \dot h - u \right) = 3 {\dot a \over a} 
                   \left( p_b - c_s{}^2 \rho_b \right) \delta ,
\end{eqnarray}
\begin{eqnarray}
\label{fluidpert2s}
       \ddot h + 2 {\dot a \over a} \dot h = {8 \pi \over m_p{}^2}
                   \left( 1 + 3 c_s{}^2 \right) \rho_b \delta ,
\end{eqnarray}
\begin{eqnarray}
\label{fluidpert3s}
       \left[ a^5 \left( \rho_b + p_b\right) u \right]_{,0} 
                   = A(A + 2) a^3 c_s{}^2 \rho_b \delta ,
\end{eqnarray}
\begin{eqnarray}
\label{fluidpert4s}
       \dot {\cal H} = {A(A + 2) \over (A - 1)(A + 3)}
                   \left[ {24\pi \over m_p{}^2} {a^2 (\rho_b + p_b) u
                   \over A(A + 2) } - \dot h \right] ,
\end{eqnarray}
\begin{eqnarray}
\label{fluidpert5s}
       \ddot h 
         & + & 6 {\dot a \over a} \dot h + {(A^2 + 2 A - 4) \over a^2} h
               + \ddot {\cal H} + 3 {\dot a\over a} \dot {\cal H} \nonumber \\
         & + & {(A^2 + 2 A - 4) \over a^2} {\cal H} 
               + {24 \pi \over m_p{}^2} \left( 1 - c_s{}^2 \right) \rho_b 
               \delta = 0 ,
\end{eqnarray}
\begin{eqnarray}
\label{fluidpert6s}
       \ddot {\cal H} + 3 {\dot a \over a} \dot {\cal H} 
                   - {A(A + 2) \over 3 a^2} {\cal H} 
                   - {A(A + 2) \over 3 a^2} h = 0 . 
\end{eqnarray}

\subsection{Scalar field as spacetime-dependent `speed of 
     sound' fluid}
\label{scalarfieldfluid}

We have shown that in the spatially flat and spatially open models the 
synchronous gauge linear perturbation equations of a fluid model with a 
given spacetime-dependent speed of propagation of `acoustic' disturbances
are identical to those of a scalar field model, Sec.\ II of Ref.\ 
\citep{Ratra1991a} and Sec.\ III.C of Ref.\ \citep{RatraPeebles1995}. 
Here we show that this result also holds in the closed model.

Defining the background energy density and pressure of the scalar
field
\begin{eqnarray}
\label{scalarrho}
     \rho_{b\Phi} = {m_p{}^2 \over 32\pi} \left[ \dot\Phi_b{}^2
                      + V(\Phi_b) \right] , 
\end{eqnarray}
\begin{eqnarray}
\label{scalarp}
     p_{b\Phi} = {m_p{}^2 \over 32\pi} \left[ \dot\Phi_b{}^2
                      - V(\Phi_b) \right] ,
\end{eqnarray}
and the fractional energy density, peculiar velocity, and `speed of
sound' of the scalar field perturbation,
\begin{eqnarray}
\label{scalarfluidpert1}
     \rho_{b\Phi} \delta_\Phi = {m_p{}^2 \over 16\pi} \left[
                                \dot\Phi_b\dot\phi + {1\over 2}
                                V'(\Phi_b)\phi \right] ,
\end{eqnarray}
\begin{eqnarray}
\label{scalarfluidpert2}
     a^2 \left( \rho_{b\Phi} + p_{b\Phi} \right) H_{ij} u^j_\Phi = 
                          - {m_p{}^2  \over 16\pi} \dot\Phi_b \partial_i
                          \phi ,
\end{eqnarray}
\begin{eqnarray}
\label{scalarfluidpert3}
     c_{s\Phi}{}^2 \rho_{b\Phi} \delta_\Phi =  {m_p{}^2 \over 16\pi}
                        \left[ \dot\Phi_b\dot\phi - {1\over 2}
                        V'(\Phi_b)\phi \right]
\end{eqnarray}
we see that the fluid stress-energy tensor, eqs.\ (\ref{fluidT00}) -- 
(\ref{fluidTij}), coincides with the scalar field stress-energy tensor, 
eqs.\ (\ref{scalarT00}) -- (\ref{scalarTij}).
It is straightforward to show that when eqs.\ (\ref{scalarrho}) and 
(\ref{scalarp}) are used in eqs.\ (\ref{fluiddensity}) -- 
(\ref{fluidfriedmann2}) these homogeneous fluid equations coincide with 
the homogeneous scalar field equations, eqs.\ (\ref{scalarfieldKG}) -- 
(\ref{scalarfieldfriedmann2}). Using the definitions of eqs.\ 
(\ref{scalarfluidpert1}) -- (\ref{scalarfluidpert3}) in the fluid spatial 
irregularity equations (\ref{fluidpert2}), (\ref{fluidpert4}) and 
(\ref{fluidpert5}), we find that they reproduce the scalar field spatial 
irregularity equations (\ref{scalarpert2}) -- (\ref{scalarpert4}). It may 
also be shown that when the definitions of eqs.\ (\ref{scalarfluidpert2}) and 
(\ref{scalarfluidpert3}) are used in eq.\ (\ref{fluidpert3}) this equation 
reduces to an identity (if the equation for the spatially homogeneous part 
of the scalar field, eq.\ (\ref{scalarfieldKG}), is satisfied). It is only 
a little bit more involved to show that the definitions (\ref{scalarrho}) 
-- (\ref{scalarfluidpert3}) imply that eq.\ (\ref{fluidpert1})
reduces to eq.\ (\ref{scalarpert1}) (the manipulations are very similar 
to those outlined at the end of Sec.\ II of Ref.\ \citep{Ratra1991a}).

\subsection{Gauge-invariant variables}
\label{gaugeinvariantvariables}

Choosing synchronous gauge does not completely fix general 
coordinate invariance --- there are four remaining time-independent
gauge symmetries. Their effect on the metric perturbation is
\begin{eqnarray}
      \delta h_{ij} (\Omega , t) = 
            & - & \left( f^0{}_{|i|j}(\Omega)
                         + f^0{}_{|j|i}(\Omega) \right)
                         \int^t {dt' \over a^2(t')} 
                                   - \omega_{i|j}(\Omega) \nonumber \\
            & - & \omega_{j|i}(\Omega) - 2 {\dot a\over a} 
                         f^0(\Omega) H_{ij}(\Omega) , 
\end{eqnarray}
where the general coordinate transformation parameters $f^0$
and $\omega_i$ are time independent.
The scalar field perturbation and the variables derived from it
transform according to
\begin{eqnarray}
     \delta\phi(\Omega , t) & = & \dot\Phi_b f^0(\Omega) , \\
     \delta \left[\delta_\Phi (\Omega , t) \right] & = &
            {\dot\rho_{b\Phi} \over \rho_{b\Phi} } f^0(\Omega) , \\
     \delta u^i_\Phi (\Omega , t) & = & - {1 \over a^2} H^{ij} f^0{}_{|j} 
            (\Omega) , \\
     \delta \left[ \left\{ c_{s\Phi}{}^2 \delta_\Phi \right\}
            (\Omega , t) \right] 
            & = & {\dot p_{b\Phi} \over \rho_{b\Phi} } f^0 (\Omega) ,
\end{eqnarray}
while the fluid variables transform, as expected, according to
\begin{eqnarray}
     \delta\left[\delta(\Omega , t) \right] & = & {\dot\rho_b \over \rho_b}
                   f^0 (\Omega) , \\
     \delta u^i (\Omega , t) & = & - {1\over a^2} H^{ij} f^0{}_{|j} 
                   (\Omega) , \\
     \delta \left[ \left\{ c_s{}^2 \delta\right\} (\Omega , t)\right]
                   & = & {\dot p_b \over \rho_b} f^0(\Omega) .
\end{eqnarray}
In spatial momentum space the scalar parts of the fields transform as
\begin{eqnarray}
   & {} &\delta{\cal H}(A,B,C,t)  =  - 2A(A+2) f^0(A,B,C) \int^t{dt'\over
              a^2(t')} \nonumber \\
             & {} & \ \ \ \ \ \ + 2\omega(A,B,C) , \\
   & {} & \delta h(A,B,C,t)  =  2A(A+2)f^0(A,B,C) \int^t{dt'\over a^2(t')} 
                           \nonumber \\ 
             & {} & \ \ \ \ \ \ - 2 \omega(A,B,C) - 6 {\dot a \over a} 
                            f^0(A,B,C) , 
\end{eqnarray}
\begin{eqnarray}
   & {} & \delta\phi(A,B,C,t) =  \dot\Phi_b f^0(A,B,C) , \\
   & {} & \delta\left[\delta(A,B,C,t)\right]  =  {\dot\rho_b\over\rho_b}
              f^0(A,B,C) , \\
   & {} & \delta\left[u(A,B,C,t)\right]  =  {A(A+2)\over a^2} f^0(A,B,C) ,\\
   & {} & \delta\left[\left\{ c_s{}^2\delta\right\}(A,B,C,t)\right]
             =  {\dot p_b \over \rho_b} f^0(A,B,C) , 
\end{eqnarray}
where $\omega_i$ and $\omega$ obey a relation like eq.\ 
(\ref{videcomposition}). 

\begin{widetext}
Following Ref.\ \citep{Ratra1991a}, it may be shown that all 
gauge-invariant information about the scalar part of the fluid perturbations 
is encoded in the gauge-invariant combinations
%\begin{widetext}
\begin{eqnarray}
\label{GIDelta}
     \Delta(A,B,C,t) = \delta(A,B,C,t)
                     + 3 {\dot a \over a} \left( {\rho_b
                     + p_b \over \rho_b}\right) {a^2 u(A,B,C,t) \over A(A+2)},
\end{eqnarray}
\begin{eqnarray}
\label{GIA}
     A(A,B,C,t)  =  \delta(A,B,C,t) 
                 - {\rho_b + p_b \over 2\rho_b}
                     \left[ h(A,B,C,t) + {\cal H}(A,B,C,t)\right]
\end{eqnarray}
(the variable $A$ should not be confused with the
spatial momentum $A$). In the scalar field model eqs.\ (\ref{GIDelta})
and (\ref{GIA}) may be rewritten, using eqs.\ (\ref{scalarrho}) -- 
(\ref{scalarfluidpert2}), as
\begin{eqnarray}
     \Delta_\Phi = { 1\over \dot\Phi_b{}^2 + V(\Phi_b)} \left[
                   2\dot\Phi_b\dot\phi + V'(\Phi_b)\phi + 6 {\dot a\over a}
                   \dot\Phi_b\phi\right] , \\
     A_\Phi = {1 \over \dot\Phi_b{}^2 + V(\Phi_b) }
              \left[ 2 \dot\Phi_b\dot\phi + V'(\Phi_b)\phi 
              - \dot\Phi_b{}^2 (h + {\cal H}) \right] . 
\end{eqnarray}
\end{widetext}

We now record the equations of motion for the fluid gauge-invariant 
variables, $\Delta$ and $A$. We have need only
for the equations in the ideal fluid model, so we set
\begin{eqnarray}
    c_s{}^2 = \nu ,
\end{eqnarray}
where $\nu$ is a numerical constant defined in eq.\ (\ref{eos}).
It is convenient to work with
\begin{eqnarray}
\label{Ddefn}
    D = { A / (\rho_b + p_b)} ,
\end{eqnarray}
instead of the variable $A$ of eq.\ (\ref{GIA}).
Using the fluid equations of motion, eqs.\ (\ref{fluiddensity}) -- 
(\ref{fluidfriedmann2})
and (\ref{fluidpert1s}) -- (\ref{fluidpert6s}), we find that $\Delta$ 
and $D$ obey
\begin{widetext}
\begin{eqnarray}
\label{DDeltaeom1}
    \dot\Delta + \left[{3\over 2} (1 - \nu) {\dot a \over a} +
                     \left\{ {(1 + 3 \nu)\over 2} + {A(A+2)\over 3}
                     \right\} {1\over a\dot a} \right]\Delta 
               = {1 \over 3} (A - 1) (A + 3) (1 + \nu) 
                     {\rho_b \over a\dot a} D ,
\end{eqnarray}
\begin{eqnarray}
\label{DDeltaeom2}
    \dot D - \left[{ A(A + 2) \over 3a \dot a} + 3(1+\nu){\dot a \over a}
                 \right] D = 
           - {A(A + 2) \over (1 + \nu) \rho_b}
                 \bigg[ \left\{ {1\over 3} + {3\over 2} {(1 + \nu)
                 \over (A-1) (A+3)} \right\} {1 \over a\dot a}  
           + {3 \over 2} { (1 + \nu) \over (A-1) (A+3)}
                 {\dot a \over a} \bigg] \Delta .
\end{eqnarray}
These equations may be combined to yield 
\begin{eqnarray}
\label{Deltaeom}
      \ddot\Delta + (2 - 3\nu){\dot a\over a}\dot\Delta 
                        + \bigg[ - {3\over 2} (1-\nu) (1+3\nu)
                                   \left({\dot a \over a}\right)^2 
                  + \left\{ - {3\over 2} (1-\nu)(1+3\nu) + 
                                \nu A(A+2)\right\} {1\over a^2} \bigg] 
                                \Delta = 0 ; 
\end{eqnarray}
\end{widetext}
a similar second order equation may be derived for the variable $D$  --- 
since we have no need for it we do not record it here.

\section{Inflation Epoch}
\label{inflationepoch}

In this section we solve the synchronous gauge equations of motion to 
derive expressions for the spatially homogeneous and inhomogeneous fields 
in the inflation epoch. 

The potential energy density for the scalar field $\Phi$ which drives 
inflation is taken to be
\begin{eqnarray}
\label{scalarfieldpotential}
     V(\Phi) = 12 h^2 [1 - \epsilon \Phi ] ,
\end{eqnarray}   
where $h^2$ is a numerical parameter
related to the inflation epoch cosmological constant 
(the parameter $h$ should not be confused with the trace
of the metric perturbation $h$) 
and $\epsilon$ is a small numerical parameter.
(These two free parameters will be constrained by
comparing our predictions to observational data.) 
The first term, $12h^2$, is large and is responsible for driving the 
expansion
of the universe during inflation, and the term proportional to $\epsilon
\Phi$ is small and is responsible for forcing the scalar field down 
the slope. This form of potential energy density is chosen so that the
leading term acts like a cosmological constant and results in closed 
de Sitter inflation while the subleading term powers a very slowly rolling 
inflaton field.

Besides the standard expansion in spatial irregularity (or the 
Newtonian gravitational constant) used to derived the usual equations
of synchronous gauge relativistic linear perturbation theory, we shall also 
make use of an expansion in the parameter $\epsilon$ to simplify the
computation, \citep{Fischleretal1985, RatraPeebles1995}. This second 
expansion assumes that $\epsilon$ is small; we shall have to check that 
this is a consistent assumption by comparing our predictions to 
observational data and verifying that the needed numerical value of 
$\epsilon$ is indeed small.

\subsection{Spatially homogeneous background fields}

We wish to determine the solutions of eqs.\ (\ref{scalarfieldKG}) -- 
(\ref{scalarfieldfriedmann2}) for the model
with the scalar field potential energy density of eq.\ 
(\ref{scalarfieldpotential}). Our ansatz for the homogeneous fields is
\begin{eqnarray}
      \Phi_b (t) & = & \Phi_{b0} (t) + \epsilon \Phi_{b1} (t) , \\
      a(t) & = & a_0 (t) \left[ 1 + \epsilon f(t) \right] , 
\end{eqnarray}
where $\Phi_{b0} (t)$, $\Phi_{b1} (t)$, $a_0(t)$ and $f(t)$ are
independent of $\epsilon$ and will be determined below.

To lowest order in $\epsilon$ eqs.\ (\ref{scalarfieldKG}) -- 
(\ref{scalarfieldfriedmann2}) are
\begin{eqnarray}
\label{scalarfieldKG0}
      \ddot \Phi_{b0} +  3 {\dot a_0 \over a_0} \dot\Phi_{b0} = 0 ,
\end{eqnarray}
\begin{eqnarray}
      \left( {\dot a_0 \over a_0} \right)^2  - {1\over 12} \dot\Phi_{b0}{}^2
            - h^2 + {1\over a_0{}^2} = 0 , 
\end{eqnarray}
\begin{eqnarray}
      {\ddot a_0 \over a_0} +  {1\over 6} \dot\Phi_{b0}{}^2 - h^2 = 0 .
\end{eqnarray}         

The first integral of eq.\ (\ref{scalarfieldKG0}) is
\begin{eqnarray}
    \dot \Phi_{b0} (t) = \dot\Phi_{b0i} \left( {a_{0i} \over a_0(t)}
    \right)^3 ,
\end{eqnarray}
where $\dot\Phi_{b0i} (a_{0i})^3$ is a constant of integration.
This solution decreases with time, because of Hubble damping, 
and we choose the constant to be
\begin{eqnarray}
     \dot \Phi_{b0i} = 0 .
\end{eqnarray}
The lowest order solution for the scalar field is then
\begin{eqnarray}
      \Phi_{b0} (t) = \Phi_{b0i} ,
\end{eqnarray}
where $\Phi_{b0i}$ is a constant of integration. The lowest order solution 
for the scale factor is 
\begin{eqnarray}
     a_0(t) = h^{-1}\, {{\rm cosh} (ht)} .
\end{eqnarray}

The first order in $\epsilon$ parts of eqs.\ (\ref{scalarfieldKG}) -- 
(\ref{scalarfieldfriedmann2}) are
\begin{eqnarray}
     \ddot \Phi_{b1} + 3 {\dot a_0 \over a_0} \dot \Phi_{b1} 
                          - 6h^2 & = & 0 , \\
     2 {\dot a_0 \over a_0} \dot f - {2f \over a_0{}^2} +
                          h^2 \Phi_{b0i} & = & 0 , \\
     \ddot f + 2 {\dot a_0 \over a_0} \dot f + h^2 \Phi_{b0i} & = & 0
\end{eqnarray}
After some work, it may be shown that the solutions of these equations are
\begin{eqnarray}
\label{Phib1}
      \Phi_{b1}(t) & = & \bar c_0 + {\bar c_1 \over 2h} 
                        \left[ { {\rm sinh} (ht) \over {\rm cosh}^2(ht) }
                        + {\rm tan}^{-1} \left\{ {\rm sinh} (ht)
                        \right\} \right] \nonumber \\
                   & {} &  + 2 \left[ {\rm ln} \left\{ {\rm cosh} (ht) \right\}
                        - {1 \over {\rm cosh}^2(ht)} \right] , 
\end{eqnarray}
\begin{eqnarray} 
      f(t) = {1 \over 2} \Phi_{b0i} - \left[ \bar c_3 h^2 + {1\over 2}
                   \Phi_{b0i} ht \right] {\rm tanh} (ht) , 
\end{eqnarray}
where $\bar c_0$, $\bar c_1$ and $\bar c_3$ are constants of integration.

\subsection{Spatial irregularities}

We shall only have need for the order $\epsilon^0$ part of $\phi$. To this 
order eq.\ (\ref{scalarpert1s}) is
\begin{eqnarray}
  \ddot \phi_0 + 3 h {\rm tanh}(ht) \dot \phi_0 + {A(A + 2) h^2 \over
   {\rm cosh}^2 (ht) } \phi_0 = 0 .
\end{eqnarray}
The solution of this equation is
\begin{eqnarray}
\label{scalarfieldpertsol}
& {} & \phi_0(A,B,C,t) =  \\ 
& {} & {h \over {\rm cosh}(ht) } \bigg[ c_+ 
                             \left\{ {\rm sinh}(ht) - i(A+1)\right\} 
              e^{-i(A+1) {\rm tan}^{-1} \left\{ {\rm sinh} (ht) \right\}} 
              \nonumber \\
& {} & \ \ \ \  + c_- \left\{ {\rm sinh} (ht) + i(A+1)\right\} 
              e^{i(A+1) {\rm tan}^{-1} \left\{ {\rm sinh} (ht) \right\}}
                        \bigg] ,  \nonumber
\end{eqnarray}
where $c_\pm$ are $A$ dependent constants of integration which will 
be determined from quantum mechanical initial conditions. We note that, to
leading order in $\epsilon$, the two solutions in this equation are 
gauge invariant.

We shall have need for the fractional energy density and peculiar 
velocity perturbations during the inflation epoch, eqs.\ 
(\ref{scalarfluidpert1}) and (\ref{scalarfluidpert2}). Using eqs.\ 
(\ref{deltadecomposition}), (\ref{videcomposition}) and 
(\ref{scalarfieldpertsol}), and the expressions of Sec.\ IV A, we have, 
to lowest order in $\epsilon$,
\begin{eqnarray}
\label{deltaphi}
    & {} &  \delta_\Phi(A,B,C,t) = \\
    & {} &  - {\epsilon \over 6 {\rm cosh}^5(ht)} \bigg[ c_+ 
            e^{-i(A+1) {\rm tan}^{-1} \left\{ {\rm sinh} (ht) \right\}}
                       \nonumber \\
    & {} & \ \ \ \ \times \bigg[A(A+2) \bigg\{ \bar c_1 + 2h {\rm sinh}(ht)
                         \left[ {\rm cosh}^2(ht) + 2\right] \bigg\} 
                      \nonumber \\
    & {} & \ \ \ \ \ \ + 6 h {\rm cosh}^4 (ht)
                     \left[ {\rm sinh}(ht) - i(A+1)\right] \bigg] \nonumber \\
    & {} & \ \ + c_- e^{i(A+1) {\rm tan}^{-1} \left\{ {\rm sinh} (ht) \right\}}
                       \nonumber \\
    & {} & \ \ \ \ \ \ \times \bigg[A(A+2) 
                       \bigg\{ \bar c_1 + 2h {\rm sinh}(ht)
                       \left[ {\rm cosh}^2(ht) + 2\right] \bigg\} 
                       \nonumber \\
    & {} & \ \ \ \ \ \ \ \ + 6 h {\rm cosh}^4 (ht)
                     \left[ {\rm sinh}(ht) + i(A+1)\right] \bigg] \bigg] ,
                        \nonumber
\end{eqnarray}
\begin{eqnarray}
    & {} & u_\Phi(A,B,C,t) =  \\
    & {} & { A (A + 2) h^3 \over \epsilon
                        \left[ \bar c_1  + 2 h {\rm sinh} (ht) \left\{
                             {\rm cosh}^2(ht) + 2 \right\} \right]} 
                        \nonumber \\ 
    & {} & \times \bigg[\left\{ {\rm sinh} (ht) - i(A+1)\right\} c_+
           e^{-i(A+1) {\rm tan}^{-1} \left\{ {\rm sinh} (ht) \right\}}
                       \nonumber \\
    & {} & \ \ \ \ + \left\{ {\rm sinh}(ht) + i(A+1)\right\} c_-
           e^{i(A+1) {\rm tan}^{-1} \left\{ {\rm sinh} (ht) \right\}} \bigg]
                       \nonumber
\end{eqnarray}
We shall also have need for expressions for the gauge-invariant 
variables $\Delta_\Phi$ and $A_\Phi$ during inflation. We find, 
\vspace{0.51cm} to leading order in $\epsilon$,
\begin{eqnarray}
\label{DeltaPhi}
 & {} &  \Delta_\Phi(A,B,C,t)  =  \\
 & {} & {\epsilon \over 6   {\rm cosh}^5(ht)} \bigg[ c_+ 
            e^{-i(A+1) {\rm tan}^{-1} \left\{ {\rm sinh} (ht) \right\}} 
                       \nonumber \\
 & {} & \ \ \times \bigg[ 3 {\rm sinh}(ht) 
                        \bigg\{ \bar c_1 + 2h {\rm sinh}(ht)
                         \left[ {\rm cosh}^2(ht) + 2\right] \bigg\} 
                         \nonumber \\
 & {} & \ \ \ \ \ \times \left[ {\rm sinh}(ht) - i(A+1)\right] 
                         \nonumber \\
 & {} & \ \ \ - A(A+2) \bigg\{ \bar c_1 + 2h {\rm sinh}(ht)
                         \left[ {\rm cosh}^2(ht) + 2\right] \bigg\} 
                      \nonumber \\
 & {} & \ \ \ - 6 h {\rm cosh}^4 (ht) \left[ {\rm sinh}(ht) - i(A+1)\right] 
                     \bigg] \nonumber \\
 & {} & \ \ \ \ + c_- e^{i(A+1) {\rm tan}^{-1} 
                     \left\{ {\rm sinh} (ht) \right\}} \nonumber \\
 & {} & \ \ \times \bigg[ 3 {\rm sinh}(ht) 
                        \bigg\{ \bar c_1 + 2h {\rm sinh}(ht)
                         \left[ {\rm cosh}^2(ht) + 2\right] \bigg\} 
                         \nonumber \\
 & {} & \ \ \ \ \ \times \left[ {\rm sinh}(ht) + i(A+1)\right] 
                         \nonumber \\
 & {} & \ \ \ - A(A+2) \bigg\{ \bar c_1 + 2h {\rm sinh}(ht)
                         \left[ {\rm cosh}^2(ht) + 2\right] \bigg\} 
                      \nonumber \\
 & {} & \ \ \ - 6 h {\rm cosh}^4 (ht) \left[ {\rm sinh}(ht) + i(A+1)\right] 
                     \bigg] \bigg] \nonumber
\end{eqnarray}
\begin{eqnarray}
\label{APhi}
     A_\Phi(A,B,C,t) = & \delta_\Phi(A,B,C,t) . 
\end{eqnarray}
where $\delta_\Phi$ is given in eq.\ (\ref{deltaphi}).

\subsection{Initial conditions and two-point correlation functions}

Conformal time $\tilde t$ is related to $t$ through
\begin{eqnarray}
    {\rm tan} {\tilde t} = {\rm sinh} (ht) .
\end{eqnarray}
In eq.\ (\ref{scalarfieldpertsol}), defining the constants $\tilde c_\pm$,
\begin{eqnarray}
    c_\pm = \pm {i \over \sqrt{2A(A+1)(A+2)}} 
            \left({16\pi \over m_p{}^2}\right)^{1/2} \tilde c_\pm , 
\end{eqnarray}
the initial conditions, Sec.\ VII of Ref.\ \citep{Ratra1985},
require that we choose (up to an irrelevant phase)
\begin{eqnarray}
     \tilde c_+ = 1  \ \  {\rm and} \ \ \tilde c_- = 0 .
\end{eqnarray}
This is equivalent to Hawking's prescription of including only regular
Euclidean field configurations \citep{Hawking1984}, and is de Sitter
invariant, see Secs.\ VI--IX of Ref.\ \citep{Ratra1985}.

In the closed de Sitter model the $a \rightarrow 0$ limit does not lie in 
the Lorentzian section \citep{Ratra1985}, unlike in the open and flat cases.  
Hawking's prescription \citep{Hawking1984} of only including field 
configurations regular on the Euclidean section does in fact correspond to 
the ground state energy of the 
rescaled scalar field inhomogeneity not diverging as $a \rightarrow 0$, 
which in this case is a pole of the Euclidean section sphere 
\citep{Ratra1985} and in fact leads to a de Sitter invariant ground state 
scalar field two-point correlation function, \citep{Ratra1985}. 

With this choice we find that the 
equal-time scalar field perturbation two-point correlation function is
\begin{widetext}
\begin{eqnarray}
    <\! {\phi_0(A,B,C,t)\phi_0^*(A',B',C',t)}\! > 
    =  \left|\phi_0(A,B,C,t)\right|^2
       \delta_{A,A'} \delta_{B,B'} \delta_{C,C'} , 
\end{eqnarray}
\begin{eqnarray}
\label{scalar2pt}
   \left|\phi_0(A,B,C,t)\right|^2 
   = {16\pi \over m_p{}^2} {1 \over 2(A+1)a^2}
       \left[ 1 + {h^2a^2 \over A(A+2)} \right] . 
\end{eqnarray}
This result coincides with eq.\ (7.13) of Ref.\ \citep{Ratra1985}. 
We note that at late time the right hand side of eq.\ (\ref{scalar2pt}) 
becomes time independent, as does the corresponding 
two-point correlation function in flat (exponentially expanding) de Sitter
spacetime (Sec.\ V of Ref.\ \citep{Ratra1985}) as well as in open
de Sitter spacetime (Sec.\ IV of Ref.\ \citep{Ratra1994} and 
Sec.\ IV.3 of Ref.\citep{RatraPeebles1995}),
however, the dependence on spatial momentum in the long wavelength limit
are quite different in the non-flat and flat cases.

This difference in the infrared behavior is also seen in the fractional
energy density perturbation two-point correlation function. We find
\begin{eqnarray}
  <\! {\delta_\Phi(A,B,C,t)\delta_\Phi^*(A',B',C',t)}\! > 
  = \left|\delta_\Phi
  (A,B,C,t)\right|^2 \delta_{A,A'} \delta_{B,B'} \delta_{C,C'} ,
\end{eqnarray}
\end{widetext}
where the fractional energy density perturbation power spectrum is
\begin{eqnarray}
   & {} &   \left|\delta_\Phi(A,B,C,t)\right|^2 =  \epsilon^2
                {16\pi \over m_p{}^2} 
                {1 \over 2A(A+1)(A+2) a^2} \nonumber \\
   & {} & \times \bigg[ (A + 1)^2 
          + \bigg[ \sqrt{h^2 a^2 - 1} + { A(A+2) \over 6 h^5 a^4} 
                \nonumber \\
   & {} & \ \ \ \ \times \left\{ \bar c_1 + 2 h \sqrt{h^2 a^2 - 1} 
                   (h^2 a^2 + 2) \right\} \bigg]^2 \bigg] ,  
\end{eqnarray}
where $\bar c_1$ is the real constant of integration in the expression 
in eq.\ (\ref{Phib1}). In the short wavelength limit the last term in 
the inner square parentheses dominates, and at late times
\begin{eqnarray}
     \left|\delta_\Phi\right|^2 \propto {A / a^4} ,
\end{eqnarray}
which is what one finds in the flat de Sitter case 
(eqn. (3.56) of Ref.\ \citep{Ratra1992}, also see Ref.\ \citep{Ratra1989}); 
this is the scale-invariant spectrum, \citep{HPYZ}. In the long wavelength 
limit the first term in the inner square parentheses dominates at late time
\begin{eqnarray}
    \left|\delta_\Phi\right|^2 \propto {1 / A} ;
\end{eqnarray}
this suggests that in the closed model the large-scale energy
density power spectrum will break away from the scale-invariant 
form and will instead behave like an $n = -1$ spectrum, like in 
the open case, see eq.\ (4.31) of Ref.\ \citep{RatraPeebles1995}.

\section {The Radiation and Matter Epochs}

In this section we solve the equations of motion to derive expressions for the 
spatially homogeneous and inhomogeneous fields in the radiation and matter
epochs.

\subsection{The radiation epoch}

In this epoch $ \nu = 1/3 = c_s{}^2$ and from 
eq.\ (\ref{fluiddensity}) $\rho_{bR} \propto a^{-4},$ or
\begin{eqnarray}
\label{rhobR}
     \rho_{bR}(t) = {3 m_p{}^2 \over 8 \pi} {h_R{}^2 \over a^4(t)} ,
\end{eqnarray}
where $h_R$ is a constant of integration determined below.  
We shall not have need for the explicit expression for $a(t).$

It suffices to derive expressions for the gauge-invariant variables $\Delta_R$ 
and $A_R$.  Defining
\begin{eqnarray}
\label{xR}
     x = {a / h_R},
\end{eqnarray}
and using eq.\ (\ref{fluidfriedmann1}) to rewrite eq.\ (\ref{Deltaeom}) in 
the radiation epoch we have
\begin{widetext}
\begin{eqnarray}
   x^2 (1 - x^2) \Delta_R''  
   - x^3 \Delta_R' +
    \left [ -2 + {1 \over 3} A(A+2) x^2 \right ] \Delta_R = 0; 
\end{eqnarray}
here a prime denotes a derivative with respect to $x$.  The solution of 
this equation is
\begin{eqnarray}
\label{DeltaRsol}
    \Delta_R(x) = c_1^{(R)} x^2 F (1+b, 1-b; 5/2; x^2) 
                + c_2^{(R)} x^{-1} F (-1/2+b, -1/2-b; -1/2; x^2) ;
\end{eqnarray}
\end{widetext}
$c_{\pm}^{(R)}$ are spatial momentum dependent constants of integration,
determined below, the $F$'s are hypergeometric functions (Chap.\ 15 of 
Ref.\ \citep{AS}), and
\begin{eqnarray}
\label{bdefn}
    b = {1 \over 2} \left ( {A(A + 2) \over 3} \right )^{1/2} .
\end{eqnarray}

From eqs.\ (\ref{Ddefn}) and (\ref{DDeltaeom1}) we have
\begin{widetext}
\begin{eqnarray}
       A_R(x) = {3 \over (A-1)(A+3)} {(1 - x^2) \over x} \Delta_R'  
              + {3 \over (A-1)(A+3)} \left [ {1 \over x^2} + 
                 { A(A+2) \over 3} \right ] \Delta_R ,  
\end{eqnarray}
\end{widetext}
so from eq.\ (\ref{DeltaRsol}) we find
\begin{eqnarray}
\label{ARsol}
   & {} &  (A-1)(A+3) A_R(x) = \\
   & {} & 3 c_1^{(R)} \bigg[ -{4\over 5} (b^2 - 1)
                x^2 ( 1 - x^2) F (2+b, 2-b; 7/2; x^2) \nonumber \\
   & {} & \ \ \ \ \ \ \ \ + \left\{ 3 + (4b^2 - 2) x^2 \right\} 
                F (1+b, 1-b; 5/2; x^2)
                \bigg] \nonumber\\
   & {} & + 3 c_2^{(R)} \bigg[ (4b^2-1) (1-x^2)x^{-1}
                 F (1/2+b, 1/2-b; 1/2; x^2) \nonumber \\
   & {} & \ \ \ \ \ \ \ \ +(4b^2+1) x^{-1} F (-1/2+b, -1/2-b; -1/2; x^2) 
          \bigg] \nonumber
\end{eqnarray}

\subsection {The matter epoch}

In this epoch $ \nu = 0 = c_s{}^2$ and from 
eq.\ (\ref{fluiddensity}) $\rho_{bM} \propto a^{-3}$, or
\begin{eqnarray}
\label{rhobM}
    \rho_{bM}(t) = {3 m_p{}^2 \over 8 \pi} {h_M{}^2 \over a^3(t)} ,
\end{eqnarray}
where $h_M$ is a constant of integration determined below.  
We shall not have need for the explicit expression for $a(t)$.

In the matter epoch eq.\ (\ref{fluidpert3s}) reduces to
\begin{eqnarray}
    {d \over dt} \left [ a^5 \rho_{bM} u_M \right ] = 0 ,
\end{eqnarray}
and we find
\begin{eqnarray}
\label{matterusol}
     u_M(t) = {c_8^{(M)} \over a^2(t)} ,
\end{eqnarray}
where $c_8^{(M)}$ is a constant of integration. In this epoch eqs.\ 
(\ref{fluidpert1s}) and (\ref{fluidpert2s}) reduce to
\begin{eqnarray}
\label{matterdeltaeq}
   \dot \delta_M - {1 \over 2} \dot h^{(M)} + u_M = 0, 
\end{eqnarray}
\begin{eqnarray}
\label{matterheq}
   \ddot h^{(M)} + 2 {\dot a \over a} \dot h^{(M)} = 
          {8 \pi \over m_p{}^2} \rho_{bM} \delta_M ,
\end{eqnarray}
where $h^{(M)}$ is the trace of the metric perturbation in the matter epoch.
Differentiating eq.\ (\ref{matterdeltaeq}) with respect to time, adding this 
result to eq.\ (\ref{matterdeltaeq}) multiplied by $2 \dot a/a$, and using 
eqs.\ (\ref{matterusol}) and (\ref{matterheq})
we find
\begin{eqnarray}
\label{matterdeltaeom}
     \ddot \delta_M + 2 {\dot a \over a} \dot \delta_M - {4 \pi \over m_p{}^2} 
     \rho_{bM} \delta_M = 0.
\end{eqnarray}
Introducing the variable $x$, Sec.\ 11.C of Ref.\ \citep{LSSU}, 
\begin{eqnarray}
\label{xM}
     x = {a / h_M{}^2} ,
\end{eqnarray}
we find that eq.\ (\ref{matterdeltaeom}) becomes
\begin{eqnarray}
     2 x^2 (1 - x) \delta_M'' + x (3 - 4 x) \delta_M'
     - 3 \delta_M = 0 ,
\end{eqnarray}
where a prime denotes a derivative with respect to $x$.  The solution of
this equation is, \citep{LSSU},
\begin{eqnarray}
\label{matterdeltasol}
      \delta_M(x) & = & c_2^{(M)} {\sqrt{1 - x} \over x^{3/2}} \\
                  & {} & + c^{(M)} \left [ - 1 + {3 \over x} - {3\sqrt{1 - x}
      \over x^{3/2}} 
      {\rm tan}^{-1} \sqrt{{x \over 1-x}} \right ] ,
      \nonumber
\end{eqnarray}
where $c_2^{(M)}$ and $c^{(M)}$ are spatial momentum dependent constants of
integration. In terms of the variable $x$ eq.\ (\ref{matterdeltaeq}) is
\begin{eqnarray}
    h^{(M)}{}' = 2 \delta_M' + 2 h_M{}^2 \sqrt{x \over 1 - x} u_M .
\end{eqnarray}
The solution of this equation is
\begin{eqnarray}
\label{matterhsol}
    h^{(M)}(x) = c_1^{(M)} + 2 \delta_M(x) - {4 c_8^{(M)} \over h_M{}^2}
                 \sqrt{1 - x \over x} ,
\end{eqnarray}
where $c_1^{(M)}$ is a spatial momentum dependent constant of integration
and $\delta_M(x)$ is given in eq.\ (\ref{matterdeltasol}).  
It is straightforward to
verify that the solutions of eqs. (\ref{matterdeltasol}) and 
(\ref{matterhsol}) satisfy eq.\ (\ref{matterheq}).
Using eq.\ (\ref{matterusol}), eq.\ (\ref{fluidpert4s}) reduces to:
\begin{widetext}
\begin{eqnarray}
    {\cal H}^{(M)}{}' = -  {A(A + 2) \over (A - 1) (A + 3)} 
                                  h^{(M)}{}'
    + {9 \over (A-1)(A+3) h_M{}^2} {c_8^{(M)} \over x^{5/2}
    \sqrt{1 - x}} .
\end{eqnarray}
The solution of this equation is
\begin{eqnarray}
\label{mattercalHsol}
    {\cal H}^{(M)} (x) = c_9^{(M)} - {A(A + 2) \over (A - 1) (A +3)}
    h^{(M)} (x) - {6 c_8^{(M)} \over (A - 1) (A + 3) h_M{}^2}
    {\sqrt{1-x} \, (2x + 1) \over x^{3/2} } , 
\end{eqnarray}
where $c_9^{(M)}$ is a spatial momentum dependent constant of integration.
In the matter epoch eqs.\ (\ref{fluidpert2s}), (\ref{fluidpert5s}) and 
(\ref{fluidpert6s}) may be combined to give
\begin{eqnarray}
    {\dot a \over a} \dot h^{(M)} 
    + {(A-1)(A+3) \over 3a^2} \left (
    h^{(M)} + {\cal H}^{(M)} \right ) + {8 \pi \over m_p{}^2}
    \rho_{bM} \delta_M = 0 . 
\end{eqnarray}
Using eqs.\ (\ref{matterdeltasol}), (\ref{matterhsol}) and 
(\ref{mattercalHsol}), we find that this equation results in
\begin{eqnarray}
    c_9^{(M)} = {3 \over (A-1)(A+3)} \left [ c_1^{(M)} - 2 c^{(M)} \right ] .
\end{eqnarray}
It may be verified that this result with eqs.\ (\ref{matterhsol}) and 
(\ref{mattercalHsol}) satisfies eq.\ (\ref{fluidpert6s}).

The matter epoch gauge-invariant variables,
$\Delta_M$ and $A_M,$ eqs.\ (\ref{GIDelta}) and (\ref{GIA}), are
\begin{eqnarray}
\label{DeltaMsol}
       \Delta_M(x) = \left \{ c_2^{(M)} + {3 c_8^{(M)} \over A(A+2)
                       h_M{}^2} \right \} {\sqrt{1 - x} \over x^{3/2}} 
       + c^{(M)} \left [ - 1 + {3 \over x} - {3 \sqrt{1 - x} \over
             x^{3/2} } {\rm tan}^{-1} \sqrt{{x \over 1 - x}} 
             \right ] , 
\end{eqnarray}
\begin{eqnarray}
\label{AMsol}
    A_M(x) & = & {A(A+2) \over (A-1)(A+3)} \left \{
                  c_2^{(M)} + {3 c_8^{(M)} \over A(A+2) h_M{}^2} \right \}
                  { \sqrt{1 - x} \over x^{3/2}} \\
    & {} & - c^{(M)} \bigg[ 1 - {3A(A+2) \over (A-1)(A+3)} 
             \left\{ {1 \over x} - {\sqrt{1 - x} \over x^{3/2}}
             {\rm tan}^{-1} \sqrt{x \over 1 - x} \right\} 
             \bigg] . \nonumber
\end{eqnarray}
\end{widetext}

\section {Joining Conditions and Expressions for the Integration Constants}

In the previous section we have derived expressions for the spatially
homogeneous and spatially inhomogeneous fields in the radiation and
matter epochs.  These solutions depend on constants of integration, and in
this section we list the equations that determine these constants of
integration and compute them. We then approximate these expressions 
for the constants of integration by discarding the contribution from 
perturbations that were inside the Hubble radius at the reheating and 
radiation-matter transitions (since we have ignored physical processes 
that are relevant on these small length scales).

As in the models of Refs.\ \citep{Ratra1992, Ratra1991b, RatraPeebles1995}, 
the constants of integration, in the radiation and matter epochs in the model 
at hand, are determined by joining conditions at the inflation-radiation 
(or reheating) transition and the radiation-matter transition.  We  
make use of the spatially homogeneous local energy density spatial 
hypersurface transition model (discussed in Refs.\ \citep{Ratra1992, 
Ratra1991a, RatraPeebles1995}), generalized to the closed FLRW model,
to derive the needed joining conditions. The resulting joining conditions 
are identical to those in Sec.\ VI A of Ref.\ \citep{RatraPeebles1995}, as 
they must be.

\subsection {Joining conditions}

In linear theory, the scalar field is identical to a spacetime-dependent 
`speed of sound' fluid (Sec. III C), so we treat both the reheating and 
radiation-matter transitions as special cases of an equation of state 
transition between two spacetime-dependent `speed of sound' fluid epochs. 
In the transition model we consider, it occurs instantaneously when the local 
energy density drops to a critical value (at different values of synchronous
gauge time $(t)$ in different parts of space).  At the transition spatial
hypersurface we require that the equation of state and `speed of sound'
change discontinuously from the value appropriate to the pretransition
fluid to that appropriate to the posttransition fluid. We consider
a transition at $t = t_{MR}$ from an $R$ fluid characterized by the variables
$\rho_{bR}$, $p_{bR}$, $c_{sR}{}^2$, to an $M$ fluid characterized by the
variables $ \rho_{bM}$, $p_{bM}$, $c_{sM}{}^2$, with a jump in the pressure at
the transition.

Since spatial gradients in the local energy density are of first order
in the perturbations, the spatially homogeneous local energy density 
spatial hypersurfaces and the synchronous gauge constant time hypersurfaces 
coincide at lowest order. We may therefore match the scale factor and the 
spatially homogeneous part of the energy density at the corresponding 
synchronous gauge constant time spatial hypersurface,
\begin{eqnarray}
\label{ajc}
     a_M(t_{MR}) =  a_R(t_{MR}) , 
\end{eqnarray}
\begin{eqnarray}
\label{rhojc}
     \rho_{bM}(t_{MR}) =  \rho_{bR}(t_{MR}).
\end{eqnarray}

Joining conditions for the inhomogeneities are derived in Sec.\ VI A of
Ref.\ \citep{RatraPeebles1995}. For our purposes here we only need 
\begin{eqnarray}
\label{Deltajc}
       \Delta_M (t_{MR}) & = & \Delta_R (t_{MR}) , 
\end{eqnarray}
\begin{eqnarray}
\label{Ajc}
      \left({A_M \over \rho_{bM} + p_{bM}}\right)(t_{MR}) & = &
          \left({A_R \over \rho_{bR} + p_{bR}}\right)(t_{MR}) .
\end{eqnarray}

\subsection{Determining the constants of integration}

Using the joining conditions for the scale factor and the background
energy density, eqs.\ (\ref{ajc}) and (\ref{rhojc}), at the two 
transitions, we have, from eqs.\ (\ref{scalarrho}), 
(\ref{scalarfieldpotential}), (\ref{rhobR}) and (\ref{rhobM}), to leading
order in $\epsilon$,
\begin{eqnarray}
      h =  {h_R / a_{R\Phi}{}^2} , \\
      h_R =  h_M \sqrt{a_{MR}} ,
\end{eqnarray}
where $a_{R\Phi}$ and $a_{MR}$ are the values of the scale factor at the 
reheating and radiation-matter transitions and $h$, $h_R$, and $h_M$ are the 
constants in eqs.\ (\ref{scalarfieldpotential}), (\ref{rhobR}) and 
(\ref{rhobM}). We note that at the reheating transition the radiation epoch 
variable $x_R$, eq.\ (\ref{xR}), is given by
\begin{eqnarray}
    x_R(t_{R\Phi}) = {a_{R\Phi} \over h_R} = {1 \over ha_{R\Phi}} ,
\end{eqnarray}
while at the radiation-matter transition the matter epoch variable
$x_M$, eq.\ (\ref{xM}), is
\begin{eqnarray}
    x_M(t_{MR}) = {a_{MR} \over h_M{}^2} = {a_{MR}{}^2 \over h_R{}^2}
           = x_R{}^2(t_{MR}) .
\end{eqnarray}

Using the joining conditions of eqs.\ (\ref{Deltajc}) and (\ref{Ajc}) 
at the reheating transition, we find, from eqs.\ (\ref{DeltaPhi}), 
(\ref{APhi}), 
(\ref{DeltaRsol}) and (\ref{ARsol}), that to leading order in $\epsilon$ 
the radiation epoch constants of integration are given by
\begin{widetext}
\begin{eqnarray}
\label{c1R} 
  & {} &  c_1^{(R)}  =  {4i\over 9\epsilon} 
                        \left( {16\pi \over m_p{}^2}\right)^{1/2} 
                        {(A-1)(A+3) \over \sqrt{2A(A+1)(A+2)}}
                        {CDE \over h_R{}^3 x_R{}^6(t_{R\Phi}) 
                        \left\{ 1 - x_R{}^2(t_{R\Phi})\right\}} \nonumber \\
  & {} & \ \ \ \ \ \ \ \ \ 
                        \times F(-1/2 + b, -1/2 -b; -1/2; x_R{}^2(t_{R\Phi})) 
                        {\rm exp} \left\{-i(A+1) {\rm tan}^{-1} 
                        \sqrt{1 - x_R{}^2(t_{R\Phi}) \over x_R{}^2(t_{R\Phi})} 
                        \right\} , 
\end{eqnarray}
\begin{eqnarray}
\label{c2R} 
  & {} &  c_2^{(R)}  =  - {4i\over 9\epsilon} 
                        \left( {16\pi \over m_p{}^2}\right)^{1/2} 
                        {(A-1)(A+3) \over \sqrt{2A(A+1)(A+2)}}
                        {CDE \over h_R{}^3 x_R{}^3(t_{R\Phi}) 
                        \left\{ 1 - x_R{}^2(t_{R\Phi})\right\}} \nonumber \\
  & {} & \ \ \ \ \ \ \ \ \ \times     F(1 + b, 1 -b; 5/2; x_R{}^2(t_{R\Phi})) 
                        {\rm exp} \left\{-i(A+1) {\rm tan}^{-1} 
                        \sqrt{1 - x_R{}^2(t_{R\Phi}) \over x_R{}^2(t_{R\Phi})} 
                        \right\} , 
\end{eqnarray}
where
\begin{eqnarray}
 & {} & C^{-1} = (4b^2 -1) x_R{}^2(t_{R\Phi}) F(1+b,1-b;5/2;x_R{}^2(t_{R\Phi}))
                 F(1/2+b,1/2-b;1/2;x_R{}^2(t_{R\Phi})) \nonumber \\
 & {} & \ \ \ \ \ \ \ \ \ \ -3  F(1+b,1-b;5/2;x_R{}^2(t_{R\Phi}))
                 F(-1/2+b,-1/2-b;-1/2;x_R{}^2(t_{R\Phi})) \nonumber \\
 & {} & \ \ \ \ \ \ \ \ \  + { 4\over 5} (b^2 - 1) x_R{}^2(t_{R\Phi}) 
                 F(-1/2+b,-1/2-b;-1/2;x_R{}^2 (t_{R\Phi})) 
                 F(2+b,2-b;7/2;x_R{}^2(t_{R\Phi})) , 
\end{eqnarray}
\begin{eqnarray}
 D^{-1/2} = \bar c_1 x_R{}^3(t_{R\Phi}) 
 + 2 h \sqrt{ 1 - x_R{}^2(t_{R\Phi})}
                \left\{1 + 2 x_R{}^2(t_{R\Phi})\right\} ,  
\end{eqnarray}
and
\begin{eqnarray}
 & {} & E  = A(A+2) \bar c_1 h^{-1} x_R{}^5(t_{R\Phi})
             + 2 A (A+2)  x_R{}^2(t_{R\Phi}) \sqrt{ 1 - x_R{}^2(t_{R\Phi})}
                \left\{1 + 2 x_R{}^2(t_{R\Phi})\right\}  \nonumber \\
 & {} & \ \ \ \ \ \ \ \ \ + 6 \left\{ \sqrt{ 1 - x_R{}^2(t_{R\Phi})} 
                - i (A+1) x_R(t_{R\Phi}) \right\} . 
\end{eqnarray} 
where $b$ is defined in eq.\ (\ref{bdefn}) and $\bar c_1$ in 
eq.\ (\ref{Phib1}). 

Using the joining conditions of eqs.\ (\ref{Deltajc}) 
and (\ref{Ajc}) at the radiation-matter transition, we find, from eqs.\ 
(\ref{DeltaRsol}), (\ref{ARsol}), (\ref{DeltaMsol}) and (\ref{AMsol}),
that the matter epoch constants of integration $c^{(M)}$ and 
\begin{eqnarray}
\label{chatMdef}
   \hat c^{(M)} \equiv c_2^{(M)} + {3c_8^{(M)} \over A(A+2) h_M{}^2} ,
\end{eqnarray}
are given by
\begin{eqnarray}
\label{cM}
   & {} & c^{(M)} =  c_1^{(R)} \bigg[ -{3 \over 5} (b^2 - 1) x_M(t_{MR})
                                      \left\{1 - x_M(t_{MR})\right\}  
                     F(2+b, 2-b; 7/2; x_M(t_{MR})) \nonumber \\
   & {} & \ \ \ \ \ \ \ \ \ \ \ \ \ \ 
                     + {1 \over 12} \left\{ 27 -(A^2+2A+18) x_M(t_{MR}) 
                     \right\}
                     F(1+b, 1-b; 5/2; x_M(t_{MR})) \bigg] \nonumber \\
   & {} & \ \ \ \ \ \ \ \ + c_2^{(R)} \bigg[ {3 \over 4} 
                     (4b^2 - 1) x_M{}^{-1/2}(t_{MR})
                                 \left\{1 - x_M(t_{MR})\right\} 
                     F(1/2+b, 1/2-b; 1/2; x_M(t_{MR})) \nonumber \\
   & {} & \ \ \ \ \ \ \ \ \ \ \ \ \ \ 
                     - {1 \over 12} (A^2+2A-9) x_M{}^{-1/2}(t_{MR}) 
                     F(-1/2+b, -1/2-b; -1/2; x_M(t_{MR})) \bigg] ,
\end{eqnarray}
\begin{eqnarray}
\label{chatM}
   & {} & \hat c^{(M)} {4 \sqrt{1 - x_M(t_{MR})} \over 3 x_M{}^{3/2}(t_{MR})} 
             = \\
   & {} & \ \ c_1^{(R)} \bigg[ {4 \over 5} (b^2 - 1)
                           \left\{1 - x_M(t_{MR})\right\}
                           \bigg\{ 3 - x_M(t_{MR})
                           - 3 \sqrt{ 1 - x_M(t_{MR}) \over x_M(t_{MR})}
                        {\rm tan}^{-1} \sqrt{ x_M(t_{MR}) \over  
                        1 - x_M(t_{MR})} \bigg\}\nonumber \\
   & {} & \ \ \ \ \ \ \ \ \ \times F(2+b, 2-b; 7/2; x_M(t_{MR}))
               - \bigg\{9/x_M(t_{MR}) - (A^2+2A+27) 
               + (A^2+2A+6)x_M(t_{MR})/9 \nonumber \\
   & {} & \ \ \ \ \ \ \ \ \  - \left\{9/x_M(t_{MR}) - (A^2+2A+18)/3 \right\}
                       \sqrt{ 1 - x_M(t_{MR}) \over x_M(t_{MR})}
                       {\rm tan}^{-1} \sqrt{ x_M(t_{MR}) 
                       \over  1- x_M(t_{MR})}\bigg\} \nonumber \\
   & {} & \ \ \ \ \ \ \ \ \ \times F(1+b, 1-b; 5/2; x_M(t_{MR})) \bigg]
                        \nonumber \\
   & {} & \ \ + c_2^{(R)} \bigg[ (4b^2 - 1) x_M{}^{-3/2}(t_{MR})
                           \left\{1 - x_M(t_{MR})\right\}
                           \bigg\{ - 3 + x_M(t_{MR}) \nonumber \\  
   & {} & \ \ \ \ \ \ \ \ \ \   + 3 \sqrt{ 1 - x_M(t_{MR}) \over x_M(t_{MR})}
                        {\rm tan}^{-1} \sqrt{ x_M(t_{MR}) \over  
                        1 - x_M(t_{MR})} \bigg\}
                        F(1/2+b, 1/2-b; 1/2; x_M(t_{MR})) \nonumber \\
   & {} & \ \ \ \ \ \ \ \ \ \  + {1 \over 9 x_M{}^{3/2}(t_{MR})}     
                 \bigg\{ 3(A^2+2A-9)
                 - (A^2+2A-21)x_M(t_{MR}) \nonumber \\
   & {} & \ \ \ \ \ \ \ \ \ \ \ -3 (A^2+2A-9)
                   \sqrt{ 1 - x_M(t_{MR}) \over x_M(t_{MR})}
                       {\rm tan}^{-1} \sqrt{ x_M(t_{MR}) 
                       \over  1- x_M(t_{MR})}\bigg\}
                       F(-1/2+b, -1/2-b; -1/2; x_M(t_{MR}))\bigg] . \nonumber
\end{eqnarray}
\end{widetext}

\subsection{Large-scale approximation}

We have ignored small-scale processes like the production of entropy at 
reheating. Our expressions are therefore only relevant for 
large-scale perturbations. From eq.\ (\ref{eigenvalue}) we see that the 
ratio of the Hubble length to a length scale which characterizes the 
perturbations is $\sqrt{A(A+2)}/(aH)$; small-scale
perturbations are those for which this ratio is $\gg 1$.
In this subsection we approximate the expressions for the constants of
integration by discarding the contribution from small-scale
perturbations at the reheating and radiation-matter transitions. 

At reheating we have, from eq.\ (\ref{xR}),
\begin{eqnarray}
    {\sqrt{A(A+2)} \over a(t_{R\Phi})H(t_{R\Phi})} = x_R(t_{R\Phi})
       \sqrt{A(A+2) \over 1 - x_R{}^2(t_{R\Phi}) } \, , 
\end{eqnarray}
so large-scale perturbations at reheating correspond to small $x_R(t_{R\Phi})$.
Expanding eqs.\ (\ref{c1R}) and (\ref{c2R}) in this limit we find for the 
radiation epoch constants of integration
\begin{widetext}
\begin{eqnarray}
\label{c1Rls}
     c_1^{(R)} & = & - {2i\over 9\epsilon} \left({16\pi \over m_p{}^2}\right)
                  ^{1/2} {(A-1)(A+3)\over \sqrt{2A(A+1)(A+2)}}
                  {e^{-i(A+1)\pi/2}\over h^2 h_R{}^3 x_R{}^6(t_{R\Phi})} \\
     & {} & \ \ \times \bigg[ 1 + (A-1) (A+3) x_R{}^2(t_{R\Phi})
                   + \left\{ {2i\over 3} A(A+1)(A+2) - {\bar c_1 \over h} 
                      \right\} x_R{}^3(t_{R\Phi})  
                    + \cdots \bigg] , \nonumber 
\end{eqnarray}
\begin{eqnarray}
\label{c2Rls}
     c_2^{(R)} & = & {2i\over 9\epsilon} \left({16\pi \over m_p{}^2}\right)
                  ^{1/2} {(A-1)(A+3)\over \sqrt{2A(A+1)(A+2)}}
                  {e^{-i(A+1)\pi/2}\over h^2 h_R{}^3 x_R{}^3(t_{R\Phi})} \\
     & {} & \ \ \times \bigg[ 1 + \left\{ {4 \over 5} A(A+2) - {21 \over 10} 
                         \right\} x_R{}^2(t_{R\Phi})
                   + \left\{ {2i\over 3} A(A+1)(A+2) - {\bar c_1 \over h} 
                      \right\} x_R{}^3(t_{R\Phi})  
                    + \cdots \bigg] ; \nonumber 
\end{eqnarray}
\end{widetext}
we note that the $\bar c_1$ dependent contribution to these expressions
are a subleading term.

At the radiation-matter transition the relevant ratio of length scales is,
from eq.\ (\ref{xM}),
\begin{eqnarray}
   {\sqrt{A(A+2)} \over a(t_{MR})H(t_{MR})} = \sqrt{ x_M(t_{MR}) A(A+2)
        \over 1 - x_M(t_{MR}) } \, ,
\end{eqnarray}
so large-scale perturbations at this transition correspond to small
$x_M(t_{MR})$. Expanding eqs.\ (\ref{cM}) and (\ref{chatM}), and using 
eqs.\ (\ref{c1Rls}) and (\ref{c2Rls}) as well as the relation $a(t_{MR}) 
\gg a(t_{R\Phi})$, we find for the matter epoch constants of integration
\begin{eqnarray}
    c^{(M)} & = & - {i\over 2 \epsilon} \left({16\pi \over m_p{}^2}\right)
                ^{1/2} {(A-1)(A+3)\over \sqrt{2A(A+1)(A+2)}} \\
    & {} & \times {e^{-i(A+1)\pi/2}\over h^2 h_R{}^3 x_R{}^6(t_{R\Phi})}
               + \cdots \nonumber ,
\end{eqnarray} 
\begin{eqnarray}
   \hat c^{(M)} = {2 \over 45} x_M{}^{5/2}(t_{MR}) c^{(M)} + \cdots . 
\end{eqnarray}

\section {Matter Epoch `Newtonian' Spatial Hypersurface and Power Spectra}

Often, theoretical expressions characterizing large-scale structure (for 
instance, the fractional mass perturbation and the peculiar velocity 
perturbation power spectra) are given in the coordinate system in which 
the time derivative of the trace of the metric perturbation has been 
removed on a given `observational' hypersurface; this is what is known as 
the instantaneously Newtonian synchronous coordinate system, Sec.\ V of 
Ref.\ \citep{Ratra1992}.  In this section we construct this instantaneously 
Newtonian coordinate system (this is a generalization to the closed 
model of the flat model construction of Sec.\ V D of Ref.\ \citep{Ratra1992} 
so we can be brief; also see Sec.\ VII A of Ref.\ \citep{RatraPeebles1995}), 
and record the power spectra of fractional energy density and
peculiar velocity perturbations in this coordinate system. In this section we 
also record the matter epoch gauge-invariant fractional energy density 
power spectrum.

\subsection{Instantaneously `Newtonian' synchronous coordinates}

The following derivation is a generalization of that of Sec.\ V D of Ref.\
\citep{Ratra1992} so we will omit technical details here.  We
choose coordinates $\hat x^\mu = (\hat t, \hat x^i)$,
\begin{eqnarray}
      \hat t & = & t - \Delta t (t_N, \vec x) , \\
      \hat x^i & = & x^i - f^i(t, \vec x) ,
\end{eqnarray}                 
which are synchronous, and require that the time derivative of the trace of the
metric perturbation, $\hat \partial_0 \hat h(\hat x)$, vanish on a spatial
hypersurface at the `observational' time $\hat t = \hat t_N$.  For the
coordinates $\hat x^\mu$ to be synchronous we must require
\begin{eqnarray}
    f^i(t, \vec x) & = & \\
    & {} &  H^{ij}(\vec x) \partial_j \Delta t (t_N, \vec x) 
              \int^t {dt' \over a^2(t')} + \omega^i(\vec x); \nonumber
\end{eqnarray}
in what follows we set $\omega^i = 0$. The fields in the two coordinate 
systems are related by
\begin{eqnarray}
      \hat \delta(\hat x) & = & \delta(x) + {\dot \rho_b(t) \over \rho_b(t)}
                              \Delta t(t_N, \vec x), \\
      \hat u^i(\hat x) & = & u^i(x) - {1 \over a^2(t)} H^{ij}(\vec x)
                           \partial_j \Delta t (t_N, \vec x) , \\
      \hat h_{ij}(\hat x) & = & h_{ij}(x) - 2 {\dot a \over a} 
                             \Delta t(t_N, \vec x) H_{ij}(\vec x) \nonumber \\ 
                       & {} &  - H_{ik}(\vec x) f^k{}_{|j}(x)
                               - H_{kj}(\vec x) f^k{}_{|i}(x) ,
\end{eqnarray}
and from the last equation, and the matter epoch equations 
(\ref{fluidfriedmann1}) and (\ref{fluidfriedmann2}), 
we have
\begin{eqnarray}
      \hat \partial_0 \hat h(\hat x) & = & \dot h(x) + \left [ {24 \pi \over
                 m_p{}^2} \rho_b(t) - {6 \over a^2(t)} \right ] 
                 \Delta t(t_N, \vec x) \nonumber \\
                 & {} & - {2 \over a^2} H^{ij}(\vec x) 
                        \Delta t_{|i |j}(t_N, \vec x) .
\end{eqnarray}
Using the matter epoch ($\nu = 0 = c_s{}^2$) fluid equations of motion in the
unbarred coordinates, Sec.\ III B, it is straightforwardly established that
when $\hat\partial_0 \hat h (\hat t_N, \hat x^k) = 0$, 
\begin{eqnarray}
       \hat \partial_0 \hat \delta (\hat t_N, \hat x^k) + \hat u^i{}_{|i}
                  (\hat t_N, \hat x^k) & = & 0 , \\
      \hat \partial_0^2 \hat \delta (\hat x) + 2 \hat H(\hat t)
                  \hat \partial_0 \hat \delta(\hat x) & = & 
                  {4 \pi \over m_p{}^2}
                  \hat \rho_b(\hat t) \hat \delta (\hat x) ;
\end{eqnarray}
these are the Newtonian matter epoch equations of motion, Secs.\ 9.B and 
10 of Ref.\ \citep{LSSU}. Comparing the second one of these to the matter 
epoch version of eq.\ (9.19) of Ref.\ \citep{LSSU}, we find that the
Newtonian gravitational potential in these coordinates, $\hat \varphi$, obeys
\begin{eqnarray}
    {\hat \nabla^2 \hat \varphi \over \hat a^2} = {4 \pi \over m_p{}^2} 
          \hat \rho_b \hat \delta .
\end{eqnarray}

In spatial momentum space, the scalar parts of the above equations 
are
\begin{widetext}
\begin{eqnarray}
   \hat \delta(A, B, C, \hat t) = 
   \delta (A, B, C, t) +
                 {\dot \rho_b(t) \over \rho_b(t)} \Delta t (A, B, C, t_N) , \\
   \hat v (A, B, C, \hat t)  = 
                 v(A, B, C, t) + {A(A+2) \over a(t)}
                 \Delta t(A, B, C, t_N) 
\end{eqnarray}
(where $v = a u$), and
\begin{eqnarray}
\label{hatpartial0h}
      \hat \partial_0 \hat h(A, B, C, \hat t)  =  \dot h(A, B, C, t) 
      + \left [ {24 \pi \over m_p{}^2} \rho_b(t) 
                 + 2 {(A - 1)(A + 3) \over
            a^2(t) } \right ] \Delta t(A, B, C, t_N) .
\end{eqnarray}
Defining the Newtonian hypersurface by requiring
\begin{eqnarray}
    \hat \partial_0 \hat h (A, B, C, \hat t_N) = 0 ,
\end{eqnarray}
we find, in the matter epoch, from eq.\ (\ref{hatpartial0h}),
\begin{eqnarray}
   \Delta t(A, B, C, t_N) = 
   - \dot h^{(M)} (A, B, C, t_N) 
          \left [ {24 \pi \over m_p{}^2} \rho_{bM}(t_N) +
          {2 (A - 1) (A + 3) \over a^2(t_N)} \right ]^{-1} .
\end{eqnarray}
\end{widetext}
Using the matter epoch solutions of Sec.\ V B we find
\begin{eqnarray}
  & {} &  \Delta t(A, B, C, t_N) =  \\
  & {} &  h_M{}^2 
             \left [ 9 + 2 (A - 1)(A + 3) x_M(t_N) \right ]^{-1} \nonumber \\
  & {} & \times \bigg [ c_2^{(M)} \{ 3 - 2 x_M(t_N) \} 
           - 2 {c_8^{(M)} \over h_M{}^2} x_M(t_N) \nonumber \\
  & {} & \ \ \ \ + c^{(M)} \bigg\{9 \sqrt{x_M(t_N) \{ 1 - x_M(t_N) \}}
         \nonumber \\ 
  & {} & \ \ \ \ \ - \{ 9 - 6 x_M(t_N) \} {\rm tan}^{-1}
             \sqrt{ x_M(t_N) \over 1 - x_M(t_N)} \bigg\} \bigg ] , \nonumber
\end{eqnarray}
where the variables and coefficients are defined in Sec.\ V B,
and we have, for the Newtonian hypersurface 
fractional energy density and peculiar velocity,
\begin{eqnarray}
 & {} &  \hat \delta_M(A, B, C, \hat t_N) =  \\
 & {} &  \left [ 9 + 2 (A - 1) (A + 3) x_M(t_N) \right ] ^{-1} \nonumber \\
 & {} & \times \bigg [ 2 A (A + 2) \hat c^{(M)} \sqrt{1 - x_M(t_N) 
                     \over x_M(t_N)} \nonumber \\
 & {} & \ \ + c^{(M)} \bigg\{ 6 A (A + 2) - 2 (A - 1)(A + 3) x_M(t_N) 
                      \nonumber \\
 & {} & \ \ \ \  - 6 A (A + 2) \sqrt{1 - x_M(t_N) \over x_M(t_N) } 
                {\rm tan}^{-1} \sqrt{ x_M(t_N) \over 1 - x_M(t_N)}
                 \bigg\} \bigg ] , \nonumber
\end{eqnarray}
\begin{eqnarray}
  & {} &  \hat v_M (A, B, C, \hat t_N) =  \\
  & {} &  A (A + 2) \left [ 9 + 2 (A - 1)(A + 3) x_M(t_N) \right ]^{-1} 
          \nonumber \\
  & {} & \times \bigg [ \hat c^{(M)} \left({3 - 2 x_M(t_N) 
                    \over x_M(t_N)}\right)
                    + c^{(M)} \bigg\{ 9 \sqrt{1 - x_M(t_N) 
                    \over x_M(t_N)}  \nonumber \\
  & {} & \ \ \ \  - \left({9 - 6 x_M(t_N) \over x_M(t_N) }\right)
                   {\rm tan}^{-1} \sqrt{ x_M(t_N) \over 1 - x_M(t_N)}
                   \bigg\} \bigg ] , \nonumber
\end{eqnarray}
where $\hat c^{(M)}$ is defined in eq.\ (\ref{chatMdef}) and the 
other expressions are defined in Sec.\ V B.

\subsection {Power spectra}

From eqs.\ (\ref{fluidfriedmann1}), (\ref{rhobM}) and (\ref{xM}) we find, 
in the matter epoch,
\begin{eqnarray}
      x_M(t) = {\Omega_0 - 1 \over \Omega_0 (1+z)} .
\end{eqnarray}

The matter fractional energy density perturbation and peculiar velocity 
perturbation equal-time two-point correlation functions are
\begin{widetext}
\begin{eqnarray}
    <\!\hat \delta_{M} (A, B, C, \hat t_N) \hat\delta_{M}^* 
            (A', B', C', \hat t_N)\!>  & = & 
   \hat P (A, \hat t_N) \delta_{A, A'} \delta_{B, B'} 
            \delta_{C, C'} , \\
   <\!\hat v_{M} (A, B, C, \hat t_N) \hat v_{M}^* (A', B', C',
            \hat t_N)\!>  & = &
            \hat P_v(A, \hat t_N) \delta_{A,  A'} \delta_{B, B'} 
            \delta_{C, C'} ,
\end{eqnarray}
where, from the results of the previous subsection, the Newtonian hypersurface
spectra are
\begin{eqnarray}
      \hat P(A, \hat t_N) & = & W_5{}^2  \left ( {W_1 \over
            c_1} \right )^2 {(A - 1)^2 (A + 3)^2 \over A (A + 1) (A + 2)}
            \left[{A(A + 2) + e_1  \over A(A + 2) + d_1}\right]^2 , \\
      \hat P_v(A, \hat t_N) & = & W_5{}^2 \left ( {W_3 \over
            c_1} \right )^2 
            {(A - 1)^2 A (A + 2) (A + 3)^2 \over 
            (A + 1) [A(A + 2) + d_1]^2 } ,
\end{eqnarray}
\end{widetext}
where we have defined
\begin{eqnarray}
      W_1 & = & {4\over 45} \sqrt{1 + \Omega_0 z_N \over \Omega_0 - 1}
              \left[ {\Omega_0 - 1 \over \Omega_0 (1+z_{MR})}\right]^{5/2} 
              \nonumber \\
          & {} & + 6 - 2 \left [ {\Omega_0 - 1 \over \Omega_0
              (1 + z_N) } \right ] \\
          & {} & - 6 \sqrt{1 + \Omega_0 z_N \over \Omega_0 - 1} \,  
          {\rm tan}^{-1} \sqrt{\Omega_0 - 1 \over 1 + \Omega_0 z_N} , 
          \nonumber \\
      W_2 & = &  6 \left [ {\Omega_0 - 1 \over \Omega_0(1 + z_N)}
           \right ] ,
\end{eqnarray}
\begin{eqnarray}
      W_3 & = & {2 \over 45} {(2 + \Omega_0 + 3 \Omega_0 z_N )
                               \over \Omega_0 - 1} 
                \left[{\Omega_0 - 1 \over \Omega_0 (1+z_{MR})}\right]^{5/2} 
                \nonumber \\
          & {} & + 9 \sqrt{1 + \Omega_0 z_N \over \Omega_0 - 1} \\
          & {} & - 3 \left [ {2 + \Omega_0 + 3 \Omega_0 z_N \over \Omega_0 - 1}
           \right ] 
          {\rm tan}^{-1} \sqrt{\Omega_0 - 1 \over 1 + \Omega_0 z_N} , 
           \nonumber
\end{eqnarray}
\begin{eqnarray}
      W_5 & = & {1 \over 2 \epsilon} 
                \left ( {16 \pi \over m_p{}^2} \right )^{1/2}
                {(1 + z_{R\Phi})^2 \over a_0} \nonumber \\
      & {} & \ \times \sqrt{\Omega_0 \over 2 (\Omega_0 - 1) (1+z_{MR})}  ,
\end{eqnarray}
\begin{eqnarray}             
     e_1 & = & {W_2 \over W_1} = {3 c_1 \over W_1}  ,\\
     c_1 & = & {2 (\Omega_0 - 1) \over \Omega_0 (1 + z_N)} , \\
     c_2 & = & 9 - 3 c_1 , \\
     d_1 & = & {c_2 \over c_1} = {(6 + 3 \Omega_0 + 9 \Omega_0 z_N) \over 2
             (\Omega_0 - 1)} ;
\end{eqnarray}
here $z_N$, $z_{MR}$, and $z_{R \Phi}$ are the redshifts of the Newtonian
hypersurface, the radiation-matter transition, and the reheating transition.
The terms dependent on $z_{MR}$ in the expressions for $W_1$ and $W_3$ are 
the contributions of the decaying solution.

We note that the matter epoch power spectrum for the gauge-invariant variable 
$\Delta_M$, eq.\ (\ref{DeltaMsol}), is
\begin{eqnarray}
    P_\Delta (A,t) = W_5{}^2 \left({W_1 \over c_1}\right)^2
         { (A - 1)^2 (A + 3)^2 \over A(A + 1)(A + 2) } , 
\end{eqnarray}
with $z_N$ in the definitions of $W_1$ and $c_1$ above replaced by $z$.
This is the generalization of the flat-space scale-invariant spectrum
\citep{HPYZ} to the closed model \citep{WhiteScott1996}. In the 
small-scale limit, which is the same as the flat-space limit, $A$ is large 
and becomes the usual flat-space Fourier wavenumber $k$ and this power 
spectrum reduces to $P_\Delta \propto k$, the standard scale-invariant expression
\citep{HPYZ}. The full closed-space power spectrum above is plotted in Fig.\
1 of Ref.\ \citep{Oobaetal2017}, where it is compared to an almost 
scale-invariant flat-space power spectrum.

\section{Conclusion}
\label{conclusion}

Using Hawking's prescription for the quantum state of the universe as
the initial conditions, we have shown that in a closed, inflating 
universe model the late-time power spectrum of gauge-invariant energy density 
inhomogeneities is not a power law. This power spectrum depends on 
wavenumber in the way expected for a generalization to the closed model 
of the standard flat-space scale-invariant power spectrum 
\citep{WhiteScott1996}. The power 
spectrum we derive appears to differ from a number of other closed 
inflation models power spectra derived assuming different (presumably
non de Sitter invariant) initial conditions.

Recent suggestions that dynamical dark energy might provide a better 
fit to the observations requires consideration of non-flat cosmological
models. It is not yet clear if non-flat $\Lambda$CDM, without dynamical dark
energy, is able to accommodate these data. Also, even if the universe is flat, 
to properly establish spatial flatness from the CMB anisotropy data 
requires use of a physically consistent non-flat cosmological model, 
such as that developed here for the positive curvature case. The power 
spectrum we have derived in this model will also be needed for a 
proper analysis of CMB anisotropy data in a mildly closed model, which 
not only remains observationally viable but might be in better accord
with the low $\ell$ CMB anisotropy observations \citep{Oobaetal2017, Oobaetal2017b}.

\acknowledgements

I thank K.\ G{\'o}rski, J.\ Ooba, J.\ Peebles, G.\ Rocha, T.\ Souradeep, 
and N.\ Sugiyama for valuable discussions. This work was supported in 
part by DOE grant DE-SC001184.


\begin{thebibliography}{99}


\bibitem{Peebles1984}
    P.~J.~E.\ Peebles, Astrophys.\ J.\ {\bf 284}, 439 (1984).

\bibitem{CosmoRev}
    B.\ Ratra and M.~S.\ Vogeley, Publ.\ Astron.\ Soc.\ Pacific {\bf 120}, 
    235 (2008) [arXiv:0706.1565];
    J.\ Martin, C.\ R.\ Physique {\bf 13}, 566 (2012) [aXiv:1205.3365];
    A.\ Joyce, L.\ Lombriser, and F.\ Schmidt, Ann.\ Rev.\ Nucl.\ Part.\ 
    Sci.\ {\bf 66}, 95 (2016) [arXiv:1601.06133], and references
    therein.

\bibitem{Adeetal2016}
   P.~A.~R.\ Ade, et al., Astron.\ Astrophys.\ {\bf 594}, A13 (2016)
   [arXiv:1502.01589], and references therein.

\bibitem{Alametal2016}
   S.\ Alam, et al., arXiv:1607.03155, and references therein.

\bibitem{Farooqetal2016}
   O.\ Farooq, F.~R.\ Madiyar, S.\ Crandall, and B.\ Ratra, Astrophys.\ J.\ 
   {\bf 835}, 26 (2017) [arXiv:1607.03537], and references therein.

\bibitem{Hzdata}
   J.\ Simon, L.\ Verde, and R.\ Jimenez, Phys.\ Rev.\ D {\bf 71}, 123001
   (2005) [arXiv:astro-ph/0412269];
   L.\ Samushia and B.\ Ratra, Astrophys.\ J.\ {\bf 650}, L5 (2006)
   [arXiv:astro-ph/0607301];
   D.\ Stern, et al., J.\ Cosmol.\ Astropart.\ Phys.\ {\bf 1002}, 008 (2010) 
   [arXiv:0907.3149];
   Y.\ Chen and B.\ Ratra, Phys.\ Lett. B {\bf 703}, 406 (2011)
   [arXiv:1106.4294];
   C.\ Zhang, H.\ Zhang, S.\ Yuan, and T.-J.\ Zhang, Res.\ Astron.\ Astrophys.\
   {\bf 14}, 1221 (2014) [arXiv:1207.4541];
   O.\ Farooq, D.\ Mania, and B. Ratra, Astrophys.\ J.\ {\bf 764}, 138
   (2013) [arXiv:1211.4253], and references therein.

\bibitem{Hztransition}
   O.\ Farooq and B. Ratra, Astrophys.\ J.\ {\bf 766}, L7 (2013) 
   [arXiv:1301.5243];
   O.\ Farooq, S.\ Crandall, and B.\ Ratra, Phys.\ Lett. B {\bf 726}, 72 
   (2013) [arXiv:1305.1957]; 
   S.\ Capozziello, O.\ Farooq, O.\ Luongo, and B. Ratra, Phys.\ Rev.\ D 
   {\bf 90}, 044016 (2014) [arXiv:1403.1421];
   M.\ Moresco, et al., J.\ Cosmol.\ Astropart.\ Phys.\ {\bf 1605}, 
   014 (2016) [arXiv:1601.01701],
   and references therein.

\bibitem{Betouleetal2014}
   M.\ Betoule, et al., Astron.\ Astrophys.\ {\bf 568}, A22 (2014)
   [arXiv:1401.4064], and references therein.

\bibitem{Pavlovetal2014}
   A.\ Pavlov, O.\ Farooq, and B. Ratra, Phys.\ Rev.\ D 
   {\bf 90}, 023006 (2014) [arXiv:1312.5285], and references therein.

\bibitem{HII}
    E.~R.\ Siegel, et al., Mon.\ Not.\ Roy.\ Astron.\ Soc.\ {\bf 356}, 1117 
   (2005) [arXiv:astro-ph/0410612];
    D.\ Mania and B. Ratra, Phys.\ Lett. B {\bf 715}, 9 (2012)
   [arXiv:1110.5626];
   R.\ Ch{\'a}vez, et al., Mon.\ Not.\ Roy.\ Astron.\ Soc.\ {\bf 462}, 
   2431 (2016) [arXiv:1607.06458], and references therein, but also 
   see J.-J.\ Wei, X.-F.\ Wu, and F.\ Melia, Mon.\ Not.\ Roy.\ Astron.\ 
   Soc.\ {\bf 463}, 1144 (2016) [arXiv:1608.02070].

\bibitem{cluster}
   L.\ Campanelli, et al., Eur.\ Phys.\ J.\ C {\bf 72}, 2218 (2012)
   [arXiv:1110.2310];
   N.\ C.\ Devi, T.\ R.\ Choudhury, and A.\ A.\ Sen, Mon.\ Not.\ Roy.\
   Astron.\ Soc.\ {\bf 432}, 1513 (2013) [arXiv:1112.0728];
   N.\ C.\ Devi, J.\ E.\ Gonzalez, and J.\ S.\ Alcaniz, J.\ Cosmol.\ 
   Astropart.\ Phys.\ {\bf 1406}, 055 (2014) [arXiv:1401.2590];
   O.\ Avsajanishvili, N.\ A.\ Arkhipova, L.\ Samushia, and 
   T.\ Kahniashvili, Eur.\ Phys.\ J.\ C {\bf 74}, 3127 (2014)
   [arXiv:1406.0407], and references therein.

\bibitem{AngSize}
   L.~I.\ Gurvits, K.~I.\ Kellermann, and S.\ Frey, Astron.\ Astrophys.\ 
   {\bf 342}, 378 (1999) [arXiv:astro-ph/9812018];
   E.~J.\ Guerra, R.~A.\ Daly, and L.\ Wan, Astrophys.\ J.\ {\bf 544}, 659 
   (2000) [arXiv:astro-ph/0006454];
   G.\ Chen and B.\ Ratra, Astrophys.\ J.\ {\bf 582}, 586 (2003) 
   [arXiv:astro-ph/0207051];
   Y.\ Chen and B.\ Ratra, Astron.\ Astrophys.\ {\bf 543}, A104 (2012)
   [arXiv:1105.5660], and references therein.

\bibitem{lookback}
    L.\ Samushia, A.\ Dev, D.\ Jain, and B.\ Ratra, Phys.\ Lett. B {\bf 693}, 
    509 (2010) [arXiv:0906.2734];
    L.~N.\ Granda, A.\ Oliveros, and W. Cardona, Mod.\ Phys.\ Lett.\ A
    {\bf 25}, 1625 (2010) [arXiv:0905:1976];
     M.~A.\ Dantas, J.~S.\ Alcaniz, D.\ Mania, and B.\ Ratra, Phys.\ Lett. B 
    {\bf 699}, 239 (2011) [arXiv:1010.0995];
    A.\ Rana, D.\ Jain, S.\ Mahajan, and A.\ Mukherjee, J.\ Cosmol.\ 
    Astropart.\ Phys.\ {\bf 1703}, 028 (2017) [arXiv:1611.07196], and 
    references therein.

\bibitem{GRB}
   L.\ Samushia and B.\ Ratra, Astrophys.\ J.\ {\bf 714}, 1347 (2010)
   [arXiv:0905.3836];
   N.\ Liang and Z.-H.\ Zhu, Res.\ Astron.\ Astrophys.\
   {\bf 11}, 497 (2011) [arXiv:1010.2681];
   M.\ Demianski, E.\ Piedipalumbo, D.\ Sawant, and L.\ Amati,
   Astron.\ Astrophys.\ {\bf 598}, A113 (2017) [arXiv:1609.09631], 
   and references therein.

\bibitem{gasmass}
   G.\ Chen and B. Ratra, Astrophys.\ J.\ {\bf 612}, L1 (2004) 
   [arXiv:astro-ph/0405636];
   L.\ Samushia, G.\ Chen, and B. Ratra, arXiv:0706.1963;
   A.~B.\ Mantz, et al., Mon.\ Not. Roy.\ Astron.\ Soc.\ {\bf 440}, 
   2077 (2014) [arXiv:1402.6212], and references therein.

\bibitem{nearfuture}
   A.\ Pavlov, L.\ Samushia, and B.\ Ratra, Astrophys.\ J.\ {\bf 760}, 19 
   (2012) [arXiv:1206.3123];
   M.\ Arabsalmani, V.\ Sahni, and T.~D.\ Saini, Phys.\ Rev.\ D {\bf 87}, 
   083001 (2013) [arXiv:1301.5779];
   L.\ Amendola, et al., arXiv:1606.00180; 
   E.\ Di Valentino, et al., arXiv:1612.00021, and references therein.

\bibitem{ChenRatra2011}
   G.\ Chen and B. Ratra, Publ.\ Astron.\ Soc.\ Pacific {\bf 123}, 1127 (2011) 
   [arXiv:1105.5206]; for earlier median statistics estimates see  
   J.~R.\ Gott, M.~S.\ Vogeley, S.\ Podariu, and B. Ratra, Astrophys.\ J.\ 
   {\bf 549}, 1 (2001) [arXiv:astro-ph/0006103];
   G.\ Chen, J.~R.\ Gott, and B. Ratra, Publ.\ Astron.\ Soc.\ Pacific 
   {\bf 115}, 1269 (2003) [arXiv:astro-ph/0308099].

\bibitem{CMBH0}
   G.\ Hinshaw, et al., Astrophys.\ J.\ Supp.\ {\bf 208}, 19 (2013) 
   [arXiv:1212.5226];
   J.~L.\ Sievers, et al., J.\ Cosmol.\ Astropart.\ Phys.\ {\bf 1310}, 
   060 (2013) [arXiv:1301.0824],
   and references therein.

\bibitem{BAOH0}
   E.\ Aubourg, et al., Phys.\ Rev.\ D {\bf 92}, 123516 (2015) 
   [arXiv:1411.1074];
   B.\ L'Huillier and A.\ Shafieloo, J.\ Cosmol.\ Astropart.\ Phys.\ 
   {\bf 1701}, 015 (2017) [arXiv:1606.06832];
   J.~L.\ Bernal, L.\ Verde, and A.~G.\ Riess, J.\ Cosmol.\ Astropart.\ Phys.\ 
   {\bf 1610}, 019 (2016) [arXiv:1607.05617];
   V.~V.\ Lukovi{\'c}, R.\ D'Agostino, and N.\ Vittorio, Astron.\ Astrophys.\
   {\bf 595}, A109 (2016) [arXiv:1607.05677], and references therein.

\bibitem{HzH0} 
   Y.\ Chen, S.\ Kumar, and B.\ Ratra, Astrophys.\ J.\ {\bf 835}, 86
   (2017) [arXiv:1606.07316].

\bibitem{Calabreseetal2012}
   E.\ Calabrese, M.\ Archidiacono, A.\ Melchiorri, and B. Ratra, 
   Phys.\ Rev.\ D {\bf 86}, 043520 (2012) [arXiv:1205.6753].

\bibitem{Riessetal2016}
   A.~G.\ Riess, et al., Astrophys.\ J.\ {\bf 826}, 56 (2016) 
   [arXiv:1604.01424];
   W.~L.\ Freedman, et al., Astrophys.\ J.\ {\bf 758}, 24 (2012) 
   [arXiv:1208.3281], and references therein.

\bibitem{Hzdynamical}
   V.\ Sahni, A.\ Shafieloo, and A.~A.\ Starobinsky, Astrophys.\ J.\ 
   {\bf 793}, L4 (2014) [arXiv:1406.2209];
   X.\ Ding, et al., Astrophys.\ J.\ {\bf 803}, L22 (2015) [arXiv:1503.04923];
   X.\ Zheng, et al., Astrophys.\ J.\ {\bf 825}, 17 (2016) [arXiv:1604.07910], 
   and references therein. 
   
\bibitem{Solaetal}
   J.\ Sol{\`a}, A.\ G{\'o}mez-Valent, and J.\ de Cruz P{\'e}rez, Astrophys.\
   J.\ {\bf 811}, L14 (2015) [arXiv:1506.05793]; Astrophys.\ J.\ {\bf 836}, 
   43 (2017) [arXiv:1602.02103]; Mod.\ Phys.\ Lett.\ A {\bf 32}, 1750054 
   (2017) [arXiv:1610.08965]; 
   J.\ Sol{\`a}, J.\ de Cruz P{\'e}rez, A.\ G{\'o}mez-Valent, and R.~C.\ Nunes,
   arXiv:1606.00450;
   J.\ Sol{\`a}, J.\ de Cruz P{\'e}rez, and A.\ G{\'o}mez-Valent,
   arXiv:1703.08218.

\bibitem{Zhangetal2017}
   Y.\ Zhang, et al., Res.\ Astron.\ Astrophys.\ {\bf 17}, 6 (2017) 
   [arXiv:1703.08293].

\bibitem{timevaryingnonflatconstraints}
   Early work on CMB anisotropy (and other) data include
   R.\ Aurich and F.\ Steiner, Mon.\ Not.\ Roy.\ Astron.\ Soc.\ {\bf 334}, 
   735 (2002) [arXiv:astro-ph/0109288]; Phys.\ Rev.\ D {\bf 67}, 123511 
   (2003) [arXiv:astro-ph/0212471]; Int.\ J.\ Mod.\ Phys.\ D {\bf 13}, 
   123 (2004)[arXiv:astro-ph/0302264]; 
   J.\ L.\ Crooks, et al., Astropart.\ Phys.\ {\bf 20}, 361 (2003) 
   [arXiv:astro-ph/0305495]; 
   K.\ Ichikawa and T.\ Takahashi,  Phys.\ Rev.\ D {\bf 73}, 083526 (2006) 
   [arXiv:astro-ph/0511821]; 
   J.\ Cosmol.\ Astropart.\ Phys.\ {\bf 0702}, 001 (2007)
   [arXiv:astro-ph/0612739]; 
   J.\ Cosmol.\ Astropart.\ Phys.\ {\bf 0804}, 027 (2008) [arXiv:0710.3995]; 
   E.\ L.\ Wright, arxiv:astro-ph/0603750; 
   K.\ Ichikawa, M.\ Kawasaki, T.\ Sekiguchi, and T.\ Takahashi, 
   J.\ Cosmol.\ Astropart.\ Phys.\ {\bf 0612}, 005 (2006) 
   [arXiv:astro-ph/0605481]; 
   G.-B.\ Zhao, et al., Phys.\ Lett.\ B {\bf 648}, 8 (2007) 
   [arXiv:astro-ph/0612728]; 
   C.\ Clarkson, M.\ Cort{\^e}s, and B. Bassett, J.\ Cosmol.\ Astropart.\ 
   Phys.\ {\bf 0708}, 011 (2007) [arXiv:astro-ph/0702670]; 
   Y.\ Wang and P.\ Mukherjee, Phys.\ Rev.\ D {\bf 76}, 103533 (2007) 
   [arXiv:astro-ph/0703780]; 
   Y.\ Gong, Q.\ Wu and A.\ Wang, Astrophys.\ J.\ {\bf 681}, 27 (2008) 
   [arXiv:0708.1817]; 
   R.\ Hlozek, M.\ Cort{\^e}s, C.\ Clarkson, and B. Bassett, Gen.\ Relativ.\ 
   Gravit.\ {\bf 40}, 285 (2008) [arXiv:0801.3847]; 
   J.-M.\ Virey, et al., J.\ Cosmol.\ Astropart.\ Phys.\ {\bf 0812}, 008 (2008)
   [arXiv:0802.4407]; 
   M.\ J.\ Mortonson, Phys.\ Rev.\ D {\bf 80}, 123504 (2009) [arXiv:0908.0346].

\bibitem{curvlimit}
   O.\ Farooq, D.\ Mania, and B. Ratra, Astrophys.\ Space Sci.\ {\bf 357}, 11
   (2015) [arXiv:1308.0834];
   D.\ Sapone, E.\ Majerotto, and S.\ Nesseris, Phys.\ Rev.\ D {\bf 90}, 023012
   (2014) [arXiv:1402.2236];
   Y.-L.\ Li, S.-Y.\ Li, T.-J.\ Zhang and T.-P.\ Li, Astrophys.\ J.\ 
   {\bf 789}, L15 (2014) [arXiv:1404.0773]; 
   R.-G.\ Cai, Z.-K. Guo, and T.\ Yang, Phys.\ Rev.\ D {\bf 93}, 043517
   (2016) [arXiv:1509.06283];
   Y.\ Chen, et al., Astrophys.\ J.\ {\bf 829}, 61 (2016) [arXiv:1603.07115];
   H.\ Yu and F.~Y.\ Wang, Astrophys.\ J.\ {\bf 828}, 85 (2016) 
   [arXiv:1605.02483];
   B.\ L'Huillier and A.\ Shafieloo, J.\ Cosmol.\ Astropart.\ Phys.\ 
   {\bf 1701}, 015 (2017) [arXiv:1606.06832];
   Z.\ Li, G.-J.\ Wang, K.\ Liao, and Z.-H.\ Zhu, Astrophys.\ J.\ {\bf 833}, 
   240 (2016) [arXiv:1611.00359];
   J.-J.\ Wei and X.-F.\ Wu, Astrophys.\ J.\ {\bf 838}, 160 (2017) 
   [arXiv:1611.00904];
   A.\ Rana, D.\ Jain, S.\ Mahajan, and A.\ Mukherjee, J.\ Cosmol.\ 
   Astropart.\ Phys.\ {\bf 1703}, 028 (2017) [arXiv:1611.07196].

\bibitem{Leonardetal2016}
   C.~D.\ Leonard, P.\ Bull, and R.\ Allison, Phys.\ Rev.\ D {\bf 94}, 023502
   (2016) [arXiv:1604.01410].

\bibitem{PeeblesRatra1988}
   P.~J.~E.\ Peebles and B.\ Ratra, Astrophys.\ J.\ {\bf 325}, L17 (1988).

\bibitem{RatraPeebles1988}
    B.\ Ratra and P.~J.~E.\ Peebles, Phys.\ Rev.\ D {\bf 37}, 3406 (1988).

\bibitem{PodariuRatra2000}
    S.\ Podariu and B.\ Ratra, Astrophys.\ J.\ {\bf 532}, 109 (2000).

\bibitem{Pavlovetal2013}
   A.\ Pavlov, S.\ Westmoreland, K.\ Saaidi, and B.\ Ratra, Phys.\ Rev.\ D 
   {\bf 88}, 123513 (2013) [arXiv:1307.7399].

\bibitem{InflationOriginal}
   D.\ Kazanas, Astrophys.\ J.\ {\bf 241}, L59 (1980);
   K.\ Sato, Mon.\ Not.\ Roy.\ Astron.\ Soc.\ {\bf 195}, 467 (1981); 
   Phys.\ Lett.\ {\bf 99B}, 66 (1981);
   A.\ H.\ Guth, Phys.\ Rev.\ D {\bf 23}, 347 (1981).
   
\bibitem{Martinetal2014}
   J.\ Martin, C.\ Ringeval, and V.\ Vennin, Phys.\ Dark Univ.\ 
   {\bf 5-6}, 75 (2014) [arXiv:1303.3787], and references therein.

\bibitem{OpenInflation}
   J.\ R.\ Gott, Nature {\bf 295}, 304 (1982); 
   B.\ Ratra and P~ J.~E.\ Peebles, Astrophys.\ J.\ {\bf 432}, L5 (1994); 
   Phys.\ Rev.\ D {\bf 52}, 1837 (1995); 
   M.\ Kamionkowski, B.\ Ratra, D.~N.\ Spergel, and N.\ Sugiyama, 
   Astrophys.\ J.\ {\bf 434}, L1 (1994) [arXiv:astro-ph/9406069]; 
   M.\ Bucher, A.~S.\ Goldhaber and N.\ Turok, Phys.\ Rev.\ D {\bf 52}, 
   3314 (1995) [arXiv:astro-ph/9411206];
   D.~H.\ Lyth and A.\ Woszczyna, Phys.\ Rev.\ D {\bf 52}, 3338 (1995)
   [arXiv:astro-ph/9501044];
    K.\ Yamamoto, M.\ Sasaki, and T.\ Tanaka, Astrophys.\ J.\ {\bf 455}, 
   412 (1995) [arXiv:astro-ph/9501109];
   K.~M.\ G{\'o}rski, B.\ Ratra, N.\ Sugiyama, and A.J.\ Banday, 
   Astrophys.\ J.\ {\bf 444}, L65 (1995) [arXiv:astro-ph/9502034];
   A.\ Linde, Phys.\ Lett.\ B {\bf 351}, 99 (1995) [arXiv:hep-th/9503097];
   A.\ Linde and A.\ Mezhlumian, Phys.\ Rev.\ D {\bf 52}, 6789 (1995) 
   [arXiv:astro-ph/9506017]; 
   K.\ Ganga, B.\ Ratra, and N.\ Sugiyama, Astrophys.\ J.\ {\bf 461}, L61 
   (1996) [arXiv:astro-ph/9512168];
   K.~M. G{\'o}rski, et al., Astrophys.\ J.\ Suppl.\ {\bf 114}, 1 (1998)
   [arXiv:astro-ph/9608054];
   S.\ Cole, D.~H.\ Weinberg, C.~S.\ Frenk and B.\ Ratra, Mon.\ Not.\ Roy.\ 
   Astron.\ Soc.\ {\bf 289}, 37 (1997) [arXiv:astro-ph/9702082];
   K.\ Ganga, et al., Astrophys.\ J.\ {\bf 484}, 517 (1997) 
   [arXiv:astro-ph/9702186];
   B.\ Ratra, et al., Astrophys.\ J.\ {\bf 517}, 549 (1999) 
   [arXiv:astro-ph/9901014].

\bibitem{Ratra1994}
    B.\ Ratra, Phys.\ Rev.\ {\bf D50}, 5252 (1994).

\bibitem{RatraPeebles1994}
    B.\ Ratra and P~ J.~E.\ Peebles, Astrophys.\ J.\ {\bf 432}, L5 (1994).

\bibitem{RatraPeebles1995}
    B.\ Ratra, and P.~J.~E.\ Peebles, Phys.\ Rev.\ D {\bf 52}, 1837 (1995).

\bibitem{LythStewart1990}
    D.~H.\ Lyth, and E.~D.\ Stewart, Phys.\ Lett.\ B {\bf 252}, 336 (1990).

\bibitem{HPYZ}
    E.~R.\ Harrison, Phys.\ Rev.\ {\bf D1}, 2726 (1970); 
    P.~J.~ E.\ Peebles and J.~T.\ Yu, Astrophys.\ J.\ {\bf 162}, 815 (1970); 
    Ya.~B.\ Zel'dovich, Mon.\ Not.\ R.\ Astron.\ Soc.\ {\bf 160}, 1P (1972).

\bibitem{Hawking1984}
    S.~W.\ Hawking, Nucl.\ Phys.\ B {\bf 239}, 257 (1984).

\bibitem{Ratra1985}
   B.\ Ratra, Phys.\ Rev.\ {\bf D31}, 1931 (1985).

\bibitem{Linde}
   A.\ Linde, Nucl.\ Phys.\ B {\bf 372}, 421 (1992);
   A.\ Linde and A.\ Mezhlumian, Phys.\ Rev.\ D {\bf 52}, 6789 (1995) 
   [arXiv:astro-ph/9506017]; 
   A.\ Linde, J.\ Cosmol.\ Astropart.\ Phys.\ {\bf 0305}, 002 (2003)
   [arXiv:astro-ph/0303245].

\bibitem{Grattonetal2002}
   S.\ Gratton, A.\ Lewis, and N.\ Turok, Phys.\ Rev.\ D {\bf 65}, 
   043513 (2002) [arXiv:astro-ph/0111012].

\bibitem{LasenbyDoran2005}
   A.\ Lasenby and C.\ Doran, Phys.\ Rev.\ D {\bf 71}, 063502 (2005) 
   [arXiv:astro-ph/0307311].

\bibitem{Ellis}
   G.~F.~R.\ Ellis, W.\ Stoeger, P.\ McEwen, and P.\ Dunsby, Gen.\ Relativ.\ 
   Gravit.\ {\bf 34}, 1445 (2002) [arXiv:gr-qc/0109023];
   G.~F.~R.\ Ellis, P.\ McEwen, W.\ Stoeger, and P.\ Dunsby, Gen.\ Relativ.\ 
   Gravit.\ {\bf 34}, 1461 (2002) [arXiv:gr-qc/0109024];
   J.-P.\ Uzan, U.\ Kirchner, and G.~F.~R.\ Ellis, Mon.\ Not.\ Roy.\ Astron.\ 
   Soc.\ {\bf 344}, L65 (2003) [arXiv:astro-ph/0302597].

\bibitem{WhiteScott1996}
   M.\ White and D.\ Scott, Astrophys.\ J.\ {\bf 459}, 415 (1996) 
   [arXiv:astro-ph/9508157].

\bibitem{closedPk}
   A.~A.\ Starobinsky, arXiv:astro-ph/9603075;
   M.\ Zaldarriaga, U.\ Seljak, and E.\ Bertschinger, Astrophys.\ J.\ 
   {\bf 494}, 491 (1998) [arXiv:astro-ph/9704265]; 
   A.\ Lewis, A.\ Challinor, and A\ Lasenby, Astrophys.\ J.\ {\bf 538}, 473
   (2000) [arXiv:astro-ph/9911177];
   J.\ Lesgourgues and T.\ Tram, J.\ Cosmol.\ Astropart.\ Phys.\ 
   {\bf 1409}, 032 (2014) [arXiv:1312.2697] .

\bibitem{PChenetal2017}
   P.\ Chen, Y.-H.\ Lin, and D.-h.\ Yeom, arXiv:1707.01471.

\bibitem{Massoetal2008}
   E.\ Mass{\'o}, S.\ Mohanty, A.\ Nautiyal and G.\ Zsembinszki, Phys.\ Rev.\ 
   D {\bf 78}, 043534 (2008) [arXiv:astro-ph/0609349].

\bibitem{Bongaetal2016}
   B.\ Bonga, B.\ Gupt, and N.\ Yokomizo, J.\ Cosmol.\ Astropart.\ Phys.\ 
   {\bf 1610}, 031 (2016) [arXiv:1605.07556].

\bibitem{Efstathiou2003}
   G.\ Efstathiou, Mon.\ Not.\ Roy.\ Astron.\ Soc.\ {\bf 343}, L95 (2003) 
   [arXiv:astro-ph/0303127].

\bibitem{Oobaetal2017}
   J.\ Ooba, B.\ Ratra, and N.\ Sugiyama, arXiv:1707:03452.

\bibitem{Oobaetal2017b}
   J.\ Ooba, B.\ Ratra, and N.\ Sugiyama, arXiv:1710:03271. 

\bibitem{Ratra1991a}
    B.\ Ratra, Phys.\ Rev.\ {\bf D43}, 3802 (1991).

\bibitem{Harmonics}
    E.\ Lifshitz, J.\ Phys.\ (USSR) {\bf 10}, 116 (1946); 
    A.\ Z.\ Dolginov, Sov.\ Phys.\ JETP {\bf 3}, 589 (1956); 
    E.\ M.\ Lifshitz and I.\ M.\ Khalatnikov, Adv.\ Phys.\ {\bf 12}, 185 
        (1963);    
    E.\ R.\ Harrison, Rev.\ Mod.\ Phys.\ {\bf 39}, 862 (1967), and 
    references therein.

\bibitem{EHTF}
    {\it Higher Transcendental Functions}, Vol.\ I, ed.\ 
    A. Erd\'elyi (McGraw-Hill, New York, 1953).
  
\bibitem{AS}
    {\it Handbook of Mathematical Functions}, eds.\ M.\ Abramowitz
    and I.\ A.\ Stegun (Dover, New York, 1972).

\bibitem{LSSU} 
    P.~J.~E.\ Peebles, {\it The Large-Scale Structure of the
    Universe} (Princeton University Press, Princeton, 1980).

\bibitem{Ratra1988}
    B.\ Ratra, Phys.\ Rev.\ {\bf D38}, 2399 (1988).
    
\bibitem{Fischleretal1985}
    W.\ Fischler, B.\ Ratra and L.\ Susskind, Nucl.\ Phys.\ B {\bf 259}, 
    730 (1985).

\bibitem{Ratra1992}
    B.\ Ratra, Phys.\ Rev.\ {\bf D45}, 1913 (1992).

\bibitem{Ratra1989}
    B.\ Ratra, Phys.\ Rev.\ {\bf D40}, 3939 (1989).

\bibitem{Ratra1991b}
    B.\ Ratra, Phys.\ Rev.\ {\bf D44}, 365 (1991).

\end{thebibliography}
\end{document}